\documentclass[english,aps,pra, twocolumn, floatfix, showpacs, superscriptaddress, notitlepage]{revtex4-1}
\usepackage{amsmath}
\usepackage{amsfonts}
\usepackage{graphicx}
\usepackage{color}
\usepackage{hyperref}
\usepackage{mathtools}
  \begin{abstract}

Realization of strong synthetic magnetic fields in driven optical lattices has enabled implementation of topological bands in cold-atom setups.
A milestone has been reached by a recent measurement of a finite Chern number based on the dynamics of incoherent bosonic atoms. 
The measurements of the quantum Hall effect in semiconductors are related to the Chern-number measurement in a cold-atom setup; however, the design and complexity of the two types of measurements are quite different. 
Motivated by these recent developments, we investigate the dynamics of weakly interacting incoherent bosons in a two-dimensional driven optical lattice exposed to an external force, which provides a direct probe of the Chern number. 
We consider a realistic driving protocol in the regime of high driving frequency and focus on the role of weak repulsive interactions.
We find that interactions lead to the redistribution of atoms over topological bands both through the conversion of interaction energy into kinetic energy during the expansion of the atomic cloud and due to an additional heating.
Remarkably, we observe that the moderate atomic repulsion facilitates the measurement by flattening the distribution of atoms in the quasimomentum space. 
Our results also show that weak interactions can suppress the contribution of some higher-order nontopological terms in favor of the topological part of the effective model.

 \end{abstract}

\begin{document}
 \title{Dynamics of weakly interacting bosons in optical lattices with flux}
 \author{Ana Hudomal}
\affiliation{Scientific Computing Laboratory, Center for the Study of Complex Systems,
Institute of Physics Belgrade, University of Belgrade, Serbia}
\author{Ivana Vasi\'c}
\affiliation{Scientific Computing Laboratory, Center for the Study of Complex Systems,
Institute of Physics Belgrade, University of Belgrade, Serbia}
\author{Hrvoje Buljan}
\affiliation{Department of Physics, Faculty of Science, University of Zagreb, Croatia}
\author{Walter Hofstetter}
\affiliation{Institut f\"ur Theoretische Physik, Johann Wolfgang Goethe-Universit\"at, Frankfurt am Main, Germany}
\author{Antun Bala\v z}
\affiliation{Scientific Computing Laboratory, Center for the Study of Complex Systems,
Institute of Physics Belgrade, University of Belgrade, Serbia}
 \maketitle

\section{Introduction}
Ultracold atoms in optical lattices provide a perfect platform for quantum simulations of various condensed-matter phenomena \cite{Bloch2008}. 
Yet, since charge-neutral atoms do not feel the Lorentz force, a big challenge in this field was realization of synthetic magnetic fields. 
After years of effort, artificial gauge potentials for neutral atoms were implemented by exploiting atomic coupling to a suitable configuration of external lasers \cite{Lin2009, Dalibard2011}.
These techniques were further extended to optical lattices, leading to the realization of strong, synthetic, magnetic fields. As a result, important condensed-matter models -- the Harper-Hofstadter \cite{Hofstadter1976} and the Haldane model \cite{Haldane1988} -- are nowadays available in cold-atom setups \cite{Aidelsburger2013, Miyake2013, Jotzu2014, Tai2017}.
The key property of these models is their nontrivial topological content. 
In the seminal TKNN paper \cite{Thouless1982} it was shown that the quantization of the Hall conductivity observed in the integer Hall effect can be directly related to the topological index of the microscopic model - the Chern number. 

Cold-atom realizations of topological models exploit periodic driving, either through laser-assisted tunneling \cite{Aidelsburger2013, Miyake2013} or by lattice shaking \cite{Jotzu2014}. Using Floquet theory \cite{Floquet1883, Grifoni1998}, a periodically driven system can be related to the time-independent effective Hamiltonian that corresponds to a relevant condensed-matter system. The mapping is known as Floquet engineering and its important features in the context of optical lattices are discussed in Refs.~\cite{Goldman2014, Goldman2015, Eckardt2015, Aidelsburger2017, Eckardt2017, Cooper2018, Sun2018, Fujiwara2018}. Because of important differences of cold-atom setups and their condensed-matter counterparts, new quench protocols for probing topological features were proposed \cite{Price2012, Dauphin2013, Bukov2014, Price2016, Mugel2017}. Following up on these studies, the deflection of an atomic cloud as a response to external force was used to experimentally measure the Chern number in a nonelectronic system for the first time \cite{Aidelsburger2015}.

While Floquet engineering is a highly flexible and powerful technique, it poses several concerns. One of the main open questions is related to the interplay of driving and interactions which can heat up the system to a featureless, infinite-temperature regime according to general considerations \cite{DAlessio2014, Bukov2015}. In particular, it is shown that an initial Bose-Einstein condensate  in a periodically driven optical lattice may become unstable due to two-body collisions \cite{Choudhury2015} or through the mechanism of parametric resonance \cite{Kennedy2015, Bukov2015, Lellouch2017, Plekhanov2017, Lellouch2018, Michon2018, Nager2018, Boulier2018}. The preparation protocol, stability and a lifetime of strongly correlated phases, expected in the regime of strong interactions under driving is a highly debated topic at the moment \cite{Bukov2015, Lelas2016, Motruk2017}. 

 In order to further explore the role of weak atomic interactions in probing topological features, here we consider the dynamics of weakly interacting incoherent bosons in a driven optical lattice exposed to an external force.  The setup that we consider includes all basic ingredients for the Chern-number measurement \cite{Dauphin2013, Aidelsburger2015} -- the Chern number of the topological band can be extracted from the center-of-mass motion of atomic cloud in the direction transverse to the applied force. We assume an ideal initial state where the lowest topological band of the effective model is almost uniformly populated. The optimal loading sequence necessary to reach this state is considered in Refs.~\cite{Ho2016, Dauphin2017}. Following the recent experimental study \cite{Aidelsburger2015}, we assume that atoms are suddenly released from the trap and exposed to a uniform force. We perform numerical simulations for the full time-dependent Hamiltonian and take into account the effects of weak repulsive interactions between atoms within the mean-field approximation. We make a comparison between the dynamics governed by the effective and time-dependent Hamiltonian and delineate the contribution of interactions to the center-of-mass response and to the overall cloud expansion dynamics.
 Our results show that interactions lead to the undesirable atomic transitions between topological bands \cite{Bilitewski2015}, but we also find that a weak atomic repulsion can facilitate the Chern-number measurements in several ways.

The paper is organized as follows. In Sec.~\ref{s:model} we describe the model and introduce a method that we apply for the description of incoherent bosons. In Sec.~\ref{s:nonint} we address the dynamics of noninteracting incoherent bosons, and then in Sec.~\ref{s:int} we address the regime of weak repulsive interactions.
Finally, we summarize our results in Sec.~\ref{s:conclusions}.
Appendixes \ref{a:heff} to \ref{a:kd} provide further details.

\section{Model and method}\label{s:model}

In this section, we first present the driven model introduced in Ref.~\cite{Aidelsburger2015},
and then derive the corresponding effective model
and discuss 
its basic characteristics.
At the end, we explain our choice of the initial state
and outline the method that we use to treat the dynamics of weakly interacting incoherent bosons.


\subsection{Effective Floquet Hamiltonian}
Interacting bosons in a two-dimensional optical lattice
can be described by the Bose-Hubbard Hamiltonian
\begin{eqnarray}\label{eq:hbh}
 \hat{H}_\mathrm{BH}&=&-J_x\sum_{l,m} \left(\hat{a}^\dagger_{l+1,m}\hat{a}_{l,m}+\hat{a}^\dagger_{l-1,m}\hat{a}_{l,m}\right)\nonumber\\
 &-&J_y\sum_{l,m} \left(\hat{a}^\dagger_{l,m+1}\hat{a}_{l,m}+\hat{a}^\dagger_{l,m-1}\hat{a}_{l,m}\right)\nonumber\\
 &+&\frac{U}{2}\sum_{l,m}\hat{n}_{l,m}\left(\hat{n}_{l,m}-1\right),
\end{eqnarray}
where  
$\hat{a}^\dagger_{l,m}$ and $\hat{a}_{l,m}$ are creation and annihilation operators that
create and annihilate a particle at the lattice site $(l,m)=la\mathbf{e}_x+ma\mathbf{e}_y$ ($a$ is the lattice constant), 
$\hat{n}_{l,m}=\hat{a}^\dagger_{l,m}\hat{a}_{l,m}$ is the number operator,
$J_x$ and $J_y$ are the hopping amplitudes along $\mathbf{e}_x$ and $\mathbf{e}_y$,
and $U$ is the on-site interaction. In the derivation of the model (\ref{eq:hbh}) we use the single-band tight-binding approximation \cite{Bloch2008}.
Although the experimental setup \cite{Aidelsburger2015} is actually three dimensional, 
with an additional confinement in the third direction, 
 our study is simplified to a two-dimensional lattice.

In order to engineer artificial gauge field 
in the experiment \cite{Aidelsburger2015},
hopping along $\mathbf{e}_x$ was at first inhibited by an additional staggered potential 
\begin{equation}\label{eq:staggered}
\hat{W}=\frac{\Delta}{2}\sum_{l,m}(-1)^l\hat{n}_{l,m}, 
\end{equation}
and then restored using resonant laser light.
The experimental setup can be described by a time-dependent Hamiltonian
\begin{equation}\label{eq:htd0}
 \tilde{H}(t)=\hat{H}_\mathrm{BH}+\hat{V}(t)+\hat{W},
\end{equation}
where $\hat{V}(t)$ is a time-dependent modulation
\begin{align}\label{eq:drive}
 \hat{V}(t)=\kappa\sum_{l,m}~\hat{n}_{l,m}\bigg[\cos\left(\frac{l\pi}{2}-\frac{\pi}{4}\right)
 \cos\left(\omega t-\frac{m\pi}{2}+\phi_0\right)\nonumber\\
 +\cos\left(\frac{l\pi}{2}+\frac{\pi}{4}\right)
 \cos\left(-\omega t-\frac{m\pi}{2}+\frac{\pi}{2}+\phi_0\right)\bigg],
\end{align}
$\kappa$ is the driving amplitude, and $\omega = \Delta$ is the resonant driving frequency. 
We set the relative phase $\phi_0$ between the optical-lattice potential and
the running waves used for laser-assisted tunneling  to $\phi_0 = \pi/4$.

\begin{figure}[!t]
\includegraphics[width=0.45\textwidth]{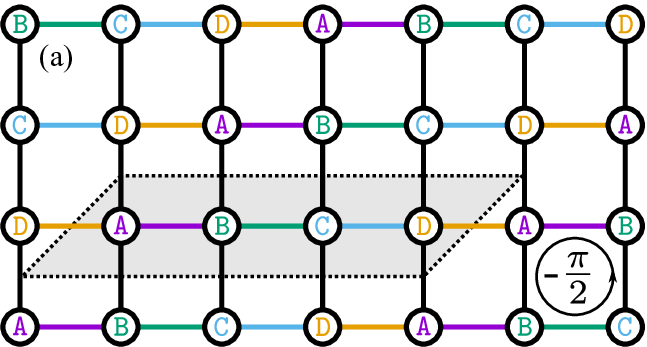} \hspace{0.5cm}
\includegraphics[width=0.45\textwidth]{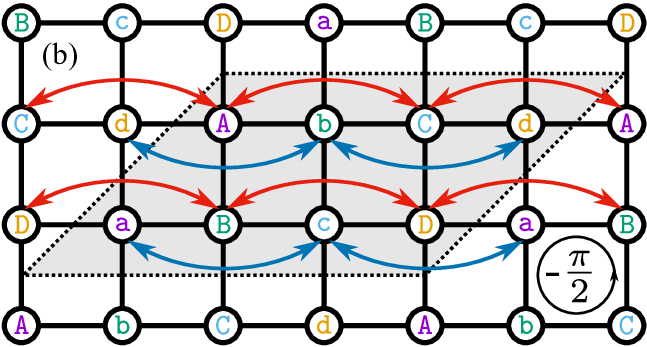}%
\caption{Schematic representation of the model. The unit cells are shaded.
(a) Effective Hamiltonian without correction, $\hat{H}_\mathrm{eff,0}$ \eqref{eq:heff0}.
Vertical links correspond to real hopping amplitudes (along $\mathbf{e}_y$ direction),
while the 
horizontal links to the right of lattice sites labeled ${\tt A}$, ${\tt B}$, ${\tt C}$, and ${\tt D}$
correspond to complex hopping amplitudes
with phases $\frac{3\pi}{4}$, $\frac{\pi}{4}$, $-\frac{\pi}{4}$, and $-\frac{3\pi}{4}$, respectively
(when hopping from left to right).
(b) Effective Hamiltonian with correction, $\hat{H}_\mathrm{eff,1}$ \eqref{eq:heff}. 
Red lines represent positive next-nearest-neighbor hopping amplitudes (connecting uppercase letters), 
while the blue lines represent negative next-nearest-neighbor hopping amplitudes (connecting lowercase letters).
Nearest-neighbor hopping amplitudes are the same as in (a). 
\label{fig:lattice}}
\end{figure}

Using Floquet theory, the time-evolution operator corresponding to the Hamiltonian \eqref{eq:htd0} can be represented as 
\begin{equation}\label{eq:floquet}
 \hat{U}(t,t_0)=\mathrm{e}^{-i\hat{W}t}\mathrm{e}^{-i\hat{K}(t)}\mathrm{e}^{-i(t-t_0)\hat{H}_\mathrm{eff}}\mathrm{e}^{i\hat{K}(t_0)}\mathrm{e}^{i\hat{W}t_0},
\end{equation}
where $\hat{H}_\mathrm{eff}$ is the full time-independent effective Hamiltonian that describes slow motion and 
$\hat{K}(t)$ is the time-periodic kick operator that describes micromotion \cite{Goldman2014,Goldman2015}.

For the moment, in this subsection we first consider the noninteracting model $U = 0$. We also assume that the driving frequency $\omega$ is the highest energy scale, but that it is still low enough that the lowest-band approximation used in deriving Eq.~(\ref{eq:hbh}) is still valid.
In the leading order of the high-frequency expansion, the effective Hamiltonian $\hat{H}_\mathrm{eff}$  is given by
\begin{align}\label{eq:heff0}
\nonumber
 \hat{H}_\mathrm{eff,0}=&J'_x\sum_{l,m}\Big[\mathrm{e}^{i\big((m-l-1)\pi/2-\pi/4\big)}\hat{a}^\dagger_{l+1,m}\hat{a}_{l,m}
+\text{h.c.}\Big]\\
&-J'_y\sum_{l,m}\left(\hat{a}^\dagger_{l,m+1}\hat{a}_{l,m}+\hat{a}^\dagger_{l,m-1}\hat{a}_{l,m}\right),
\end{align} 
where the renormalized hopping amplitudes are $J'_x=\frac{J_x\kappa}{\sqrt{2}\omega}=J_y$ and $J'_y=J_y\Big(1-\frac{1}{2}\frac{\kappa^2}{\omega^2}\Big)$.
A schematic representation of this model is presented in Fig.~\ref{fig:lattice}(a).
The unit cell is shaded and the full lattice is spanned by the vectors $\mathbf{R}_1 = (4, 0)$ and $\mathbf{R}_2 = (1, 1)$.
Particle hopping around a plaquette in the counterclockwise direction acquires a complex phase $-\frac{\pi}{2}$ and the model is equivalent to the Harper-Hofstadter Hamiltonian \cite{Hofstadter1976} for the case $\alpha=1/4$ \cite{Hofstadter1976}. The explicit form of the kick operator $\hat{K}(t)$ from Eq.~(\ref{eq:htd0}) is given in Appendix \ref{a:heff}.

Following Refs.~\cite{Goldman2014, Goldman2015}, we find that additional corrections of the order $J_x^2/\omega$ contribute to the system's dynamics and we introduce another approximation for the effective Hamiltonian
\begin{align}\label{eq:heff}
 &\hat{H}_\mathrm{eff, 1}=\hat{H}_\mathrm{eff,0}\nonumber\\
 &+\frac{J^2_x}{\omega}\sum_{l,m}(-1)^l\Big(2\hat{a}^\dagger_{l,m}\hat{a}_{l,m}
 +\hat{a}^\dagger_{l+2,m}\hat{a}_{l,m}+\hat{a}^\dagger_{l-2,m}\hat{a}_{l,m}\Big).
\end{align}
The derivation of Hamiltonian (\ref{eq:heff}) is given in Appendix \ref{a:heff} and its schematic representation is given in  Fig.~\ref{fig:lattice}(b).
The $J_x^2/\omega$ correction introduces next-nearest-neighbor hopping along $x$ direction with opposite signs for lattice sites with either even or odd $x$-coordinate $l$.
This term does not change the total complex phase per plaquette, 
but the unit cell is now doubled and thus the first Brillouin zone is halved. A similar term was engineered on purpose in order to implement the Haldane model \cite{Jotzu2014}.

In the next subsection we investigate properties of energy bands of both effective Hamiltonians, $\hat{H}_\mathrm{eff,0}$ and $\hat{H}_\mathrm{eff, 1}$.
We use the units where $\hbar=1$ and $a=1$.
Unless otherwise stated, 
we set the parameters to the following values:
lattice size $100\times100$ sites,
hopping amplitudes $J'_x=J_y=1\equiv J$, and
the driving amplitude $\kappa=0.58\;\omega$.
This value of the driving amplitude was chosen to be the same as in the experiment \cite{Aidelsburger2015}.
In order to set the renormalized hopping amplitude along $\mathbf{e}_x$ to $J'_x=1$, the initial hopping amplitude has to be 
$J_x=\sqrt{2}\omega/\kappa=2.44$,
and the correction term is therefore proportional to $J_x^2/\omega=5.95/\omega$,
so it cannot be safely neglected unless the driving frequency is very high.

\subsection{Band structure}\label{s:bands}
\begin{figure*}[!th]
\includegraphics[width=0.32\textwidth]{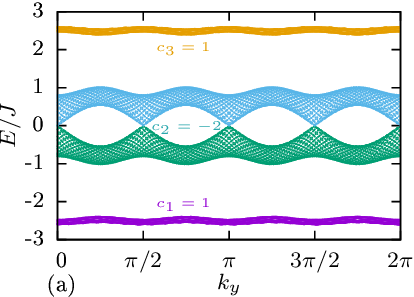}
\includegraphics[width=0.32\textwidth]{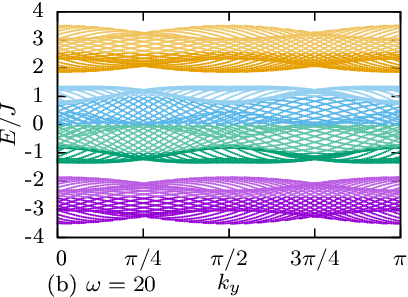}
\includegraphics[width=0.32\textwidth]{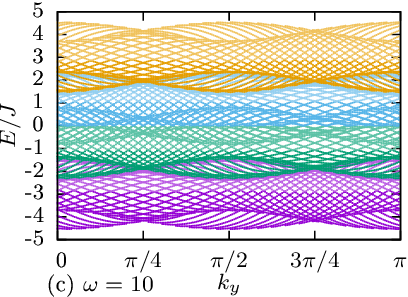}
\caption{Energy bands of the effective Hamiltonians.
(a) $\hat{\mathcal{H}}_\mathrm{eff,0}(\mathbf{k})$ Eq.~\eqref{eq:hk0}, which is without the $J_x^2/\omega$ correction term.
(b) $\hat{\mathcal{H}}_\mathrm{eff,1}(\mathbf{k})$ Eq.~\eqref{eq:hk}, which includes the correction term.
Driving frequency $\omega=20$; gaps are open.
(c) Same as (b), but with $\omega=10$. Gaps are closed. \label{fig:bands}}
\end{figure*}

Momentum-space representations of the effective Hamiltonians $\hat{H}_\mathrm{eff,0}$ and $\hat{H}_\mathrm{eff, 1}$,
denoted by $\hat{\mathcal{H}}_\mathrm{eff,0}(\mathbf{k})$ and $\hat{\mathcal{H}}_\mathrm{eff,1}(\mathbf{k})$, respectively, 
are derived in Appendix \ref{a:hk}.
Band structures for the effective Hamiltonian $\hat{\mathcal{H}}_\mathrm{eff,0}$ without the $J_x^2/\omega$ correction, Eq. \eqref{eq:hk0},
as well as for the effective Hamiltonian $\hat{\mathcal{H}}_\mathrm{eff,1}$ including the correction term, Eq. \eqref{eq:hk}, 
are shown in Fig.~\ref{fig:bands} for the two values of driving frequencies $\omega=20$ and $\omega=10$.

The Hamiltonian $\hat{H}_\mathrm{eff,0}$ is the Harper-Hofstadter Hamiltonian for the flux $\alpha=1/4$.
It has four energy bands, where the middle two bands touch at $E=0$ and can therefore be regarded as a single band;
see Fig.~\ref{fig:bands}(a). The topological content of these bands is characterized by the topological index called the Chern number.
The Chern number is the integral of the Berry curvature \cite{Berry1984} over the first Brillouin zone
divided by $2\pi$, 
\begin{equation}\label{eq:cn}
 c_n=\frac{1}{2\pi}\int_\mathrm{FBZ}\mathbf{\Omega}_n(\mathbf{k})\cdot d\mathbf{S},
\end{equation}
where $n$ denotes the band number 
and the Berry curvature is $\mathbf{\Omega}_n(\mathbf{k}) = i\nabla_\mathbf{k}\times\langle u_n(\mathbf{k})\lvert \nabla_\mathbf{k}\lvert u_n(\mathbf{k})\rangle$,
expressed in terms of eigenstates of the effective Hamiltonian $\lvert u_n(\mathbf{k})\rangle$.
The Chern numbers of the three well-separated bands are $c_1=1$, $c_2=-2$, and $c_3=1$.

Because the correction from Eq.~(\ref{eq:heff}) includes next-nearest-neighbor hopping terms, the elementary cell in real space is doubled
[see Fig.~\ref{fig:lattice}(b)] 
and, as a consequence, the first Brillouin zone for the Hamiltonian $\hat{\mathcal{H}}_\mathrm{eff,1}$ is 
reduced by a factor of $2$
compared to $\hat{\mathcal{H}}_\mathrm{eff, 0}$.
There are now eight lattice sites in the unit cell and eight energy bands, 
but the number of gaps depends on the driving frequency. 
The new bands touch in pairs, in such a way that there are always 
maximally three well-separated bands.
When the driving frequency is high enough, the correction is small and the gaps between the three bands remain open; see Fig.~\ref{fig:bands}(b). 
The original band structure of $\hat{\mathcal{H}}_\mathrm{eff,0}$ is recovered in the limit $\omega\rightarrow\infty$.
The Berry curvature and the Chern number can be calculated using the efficient method presented in Ref.~\cite{Fukui2005}.
Our calculations confirm that the Chern numbers of $\hat{\mathcal{H}}_\mathrm{eff, 1}$ are equal to those
of $\hat{\mathcal{H}}_\mathrm{eff,0}$ ($c_1=1$, $c_2=-2$, and $c_3=1$), as long as the gaps between the energy bands are open.
The gaps close when the driving frequency is too low, see Fig.~\ref{fig:bands}(c),
and the Chern numbers of the subbands 
can no longer be properly defined.

\subsection{Dynamics of incoherent bosons}\label{s:incoherent}

We need to take into account a contribution of weak, repulsive interactions. Full numerical simulations of an interacting many-body problem are computationally demanding, so we need a reasonable, numerically tractable approximation. To this end we will use the classical field method \cite{Kagan1997}, which belongs to a broader class of truncated Wigner approaches \cite{Polkovnikov2010}.
This method is similar to the approach used to treat incoherent light in instantaneous media \cite{Buljan2004,Cohen2006}, known in optics as the modal theory. 

The underlying idea of the method is to represent the initial state 
as an incoherent mixture of coherent states $\lvert \psi\rangle$, $\hat{a}_{l, m}\lvert \psi\rangle = \psi_{l, m} \lvert \psi\rangle$ \cite{Kagan1997}. 
This is explained in more detail in Appendix \ref{a:ib}.
In our study, we sample 
initial configurations of these coherent states with 
\begin{equation}\label{eq:initial}
 \lvert\psi(t = 0)\rangle=\sum_{k = 1}^{N_m}\mathrm{e}^{i\phi_k}\lvert k\rangle,
\end{equation}
where $\phi_k \in [0, 2\pi)$ are random phases and the states $|k\rangle$ correspond closely to the lowest-band eigenstates of $\hat{H}_{\text{eff}}$. 
Each of $N_{\text{samples}}$ initial states is time evolved and physical variables can be extracted by averaging over an ensemble of different initial conditions.

The time evolution of each of these coherent states is governed by
\begin{eqnarray}\label{eq:gp}
i\frac{d\psi_{l,m}(t)}{dt}&=&\sum_{ij} H_{lm, ij}(t)\psi_{i, j}(t) -F\, m\, \psi_{l, m}(t) \nonumber\\&+&U |\psi_{l, m}(t)|^2\psi_{l, m}(t),
\end{eqnarray}
where $H_{lm, ij}(t) = \langle l,m|\hat{H}(t)|i,j\rangle$ are matrix elements of $\hat{H}(t)$ from Eq.~(\ref{eq:htd0}), $F$ is the external force, and interactions $U$ contribute with the last, nonlinear term. 
Formally, Eq.~(\ref{eq:gp}) takes the form of the  Gross-Pitaevskii equation \cite{Giorgini1999, GPbook, GPbook2}.
The performances and limitations of the method are discussed and reviewed in Ref.~\cite{IBbook}.

For comparison, we also consider the related time evolution governed by the effective Hamiltonian
\begin{eqnarray}\label{eq:gp2}
i\frac{d\psi_{l,m}(t)}{dt}&=&\sum_{ij} h^{\text{eff}}_{lm, ij}\psi_{i, j}(t) -F\, m\, \psi_{l, m}(t) \nonumber\\&+&U |\psi_{l, m}(t)|^2\psi_{l, m}(t),
\end{eqnarray}
where $h^{\text{eff}}_{lm, ij} = \langle l,m|\hat{h}^{\text{eff}}|i,j\rangle$,
with $\hat{h}^{\text{eff}}$ being either $\hat{H}_{\text{eff},0}$ from Eq.~(\ref{eq:heff0}), or $\hat{H}_{\text{eff},1}$ from Eq.~(\ref{eq:heff}). Equation \eqref{eq:gp2} should be considered only as a tentative description of the system: the mapping between $\hat{H}(t)$ and $\hat{H}_{\text{eff}}$ is strictly valid only in the noninteracting regime and the interaction term may introduce complex, nonlocal, higher-order corrections \cite{DAlessio2014}. 
However, we expect their contribution to be small in the limit $U\rightarrow0$, and for time scales which are not too long \cite{Mori2016,Kuwahara2016,Abanin2017a,Abanin2017b}.

In the following we use $N_m = 300$  modes and accommodate $N_p = 300$ particles per mode, so in total in the simulations we have $N = N_mN_p = 90,000$ bosons. Typical densities in real space are up to $100$ particles per site and we choose the values of $U$ in the range $U\in [0, 0.05]$.
Other parameters: $J'_x = J_y = 1$, $\kappa/\omega = 0.58$, $\omega = 10,20$, and $F=0.25J/a$.
The correction terms are non-negligible in this frequency range.
In practice, we first numerically diagonalize the Hamiltonian (\ref{eq:hini}) from Appendix \ref{a:ib} and set our parameters in such a way that the lowest $N_m$ modes have high overlap with the lowest band of the effective model. In the next step, we sample initial configurations (\ref{eq:initial}). 
For each of $N_{\text{samples}} = 1,000$  sets of initial conditions we then time evolve Eq.~(\ref{eq:gp}) and extract quantities of interest by averaging over resulting trajectories. 
This value of $N_{\text{samples}}$ is chosen to be high enough, so that the fluctuations are weak.
We present and discuss results of our numerical simulations in the following sections.

\section{Noninteracting case}\label{s:nonint}

\begin{figure}[!t]
\includegraphics[width=0.475\textwidth]{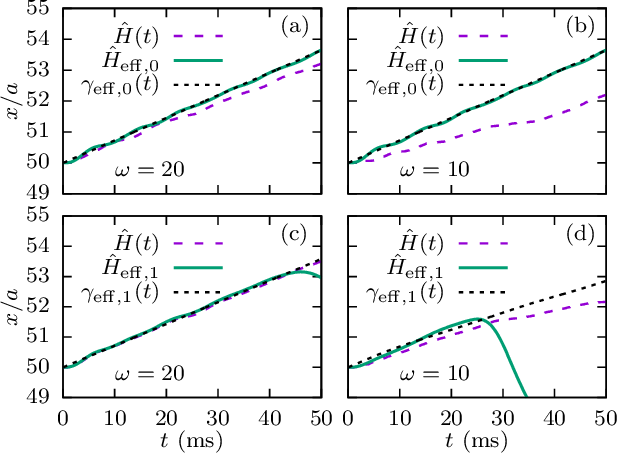}
\caption{Anomalous drift $x(t)$.
Dashed purple lines: numerical simulations using the time-dependent Hamiltonian $\hat{H}(t)$ 
(cases 2 and 4 from Table~\ref{tab:cases}).
Solid green lines: effective Hamiltonians $\hat{H}_\mathrm{eff,1}$ (c) and (d) and $\hat{H}_\mathrm{eff,0}$ (a) and (b) (cases 1 and 3).
Dotted black lines: theoretical prediction \eqref{eq:tp} from 
$\gamma_\mathrm{eff,1}(t)$ or $\gamma_\mathrm{eff,0}(t)$. 
(a) 
Initial states and band populations obtained using the
effective Hamiltonian $\hat{H}_\mathrm{eff,0}$ without the correction (cases 3 and 4).
Driving frequency $\omega=20$.
(b) $\omega=10$.
(c) Hamiltonian $\hat{H}_\mathrm{eff,1}$ with the $J_x^2/\omega$ correction (cases 1 and 2). 
 Driving frequency $\omega=20$.
(d) $\omega=10$.
\label{fig:drift}}
\end{figure}

We start by addressing the dynamics of noninteracting bosons. In this case we set $U = 0$ in Eq.~(\ref{eq:gp}) and numerically solve the single-particle Schr\"odinger equation without further approximations. Our aim is to numerically validate and compare the two approximations, Eqs.~(\ref{eq:heff0}) and (\ref{eq:heff}), for the effective Hamiltonian. To this purpose, we juxtapose results of the two approximative schemes with the numerically exact results obtained by considering the full time evolution governed by $\hat{H}(t)$. 
For clarity, the four different time evolutions that we consider in this section are summarized in Table~\ref{tab:cases}.
We calculate the center-of-mass position $x(t)$ and plot the results in Fig.~\ref{fig:drift}. In this way we also find the regime of microscopic parameters where the Chern-number measurement can be optimally performed.

First, we consider the basic Harper-Hofstadter Hamiltonian (\ref{eq:heff0}) and select the occupied modes $\lvert k \rangle $ of  the initial state (\ref{eq:rhoini}) as eigenstates of the model from Eq.~(\ref{eq:initial}) for $\hat{h}_{\text{eff}} = \hat{H}_{\text{eff},0}$. As explained in the previous section, at the initial moment $t_0=0$, the confinement is turned off and the force $\mathbf{F}=-F\mathbf{e}_y$ is turned on. As a consequence of the applied external force and the nonzero Chern number of the lowest band of the model (\ref{eq:heff0}), 
the particles exhibit an anomalous velocity in the direction perpendicular to the force \cite{Xiao2010}.
In the ideal case, when the lowest band is fully populated, 
the theoretical prediction for the center-of-mass position in the $\mathbf{e}_x$ direction is \cite{Aidelsburger2015}
\begin{equation}
x(t)=x(t_0)+c_1\frac{2Fa^2}{\pi\hbar}t,
\end{equation}
where $c_1 = 1$ is the Chern number (\ref{eq:cn}) of the lowest band. However, even in the ideal case, due to the sudden quench of the linear potential, a fraction of particles is transferred to the higher bands. To take this effect into account, the authors of Ref.~\cite{Aidelsburger2015} introduced a filling factor $\gamma(t)$
\begin{equation}\label{eq:gamma}
 \gamma(t)=\eta_1(t)-\eta_2(t)+\eta_3(t),
\end{equation}
where $\eta_i(t)$ are populations of different bands of Hamiltionian (\ref{eq:heff0}) from Eq.~(\ref{eq:eta}) in Appendix \ref{a:ib} and the plus and minus signs in Eq.~\eqref{eq:gamma} 
are defined according to the Chern numbers $c_1=1$, $c_2=-2$, and $c_3=1$. 
The final theoretical prediction is then \cite{Aidelsburger2015}
\begin{equation}\label{eq:tp}
x(t)=x(t_0)+c_1\frac{2Fa^2}{\pi\hbar}\int_0^t \gamma(t')dt'. 
\end{equation}
\begin{table}[!t]

\begin{tabular}{| c | c | c | c |}
\hline
 case & initial state & band populations & evolution\\
\hline
 1 & $\hat{H}_\mathrm{eff,1}$ & $\hat{H}_\mathrm{eff,1}$ & $\hat{H}_\mathrm{eff,1}$\\
 2 & $\hat{H}_\mathrm{eff,1}$ & $\hat{H}_\mathrm{eff,1}$ & $\hat{H}(t)$\\
 3 & $\hat{H}_\mathrm{eff,0}$ & $\hat{H}_\mathrm{eff,0}$ & $\hat{H}_\mathrm{eff,0}$\\
 4 & $\hat{H}_\mathrm{eff,0}$ & $\hat{H}_\mathrm{eff,0}$ & $\hat{H}(t)$ \\
 \hline
\end{tabular}
\caption{Four different cases: the same effective Hamiltonian is always used for the initial state
and band definitions, either with or without the correction.
The evolution is governed either by the time-dependent Hamiltonian
or by the same effective Hamiltonian as the one that was used for the initial state and calculation of band populations.
\label{tab:cases}}
\end{table}

In Fig.~\ref{fig:drift}(a) we consider the anomalous drift for a high value of the driving frequency $\omega = 20$, where we expect the expansion in $1/\omega$ to be reliable. We find an excellent agreement between the prediction (\ref{eq:tp}) (dotted black line) and numerical calculation based on $\hat{H}_{\text{eff},0}$ (solid green line).
However, some deviations between the full numerical results (dashed purple line) and the results of the approximation scheme (solid green line) are clearly visible. These deviations are even more pronounced for $\omega = 10$, Fig.~\ref{fig:drift}(b).

Now we turn to the effective model (\ref{eq:heff}). In this case we select the modes of the initial state as eigenstates of Eq.~(\ref{eq:initial}) for $\hat{h}_{\text{eff}} = \hat{H}_{\text{eff,1}}$. Moreover, we also consider band populations (\ref{eq:eta}) of the same model. In the case when $\omega=20$, Fig.~\ref{fig:drift}(c), the anomalous drift obtained using the effective Hamiltonian \eqref{eq:heff} (solid green line) closely follows the theoretical prediction \eqref{eq:tp}.
Moreover, from the same figure we can see that the effective Hamiltonian $\hat{H}_\mathrm{eff,1}$ 
reproduces the behavior of the time-dependent Hamiltonian very well. 
All three curves almost overlap for intermediate times ($5-40~\mathrm{ms}$); see Fig.~\ref{fig:drift}(c).
We attribute the long-time ($>45~\mathrm{ms}$) deviations to the finite-size effects introduced by the next-nearest-neighbor hopping terms, 
which cause the atomic cloud to reach the edge of the lattice faster. This effect is explained in more detail in Sec. \ref{s:density}.

For a lower driving frequency $\omega=10$, the effective and the time-dependent Hamiltonians do not agree so well anymore; see Fig.~\ref{fig:drift}(d).
The finite-size effects can be observed even earlier in this case (around $25~\mathrm{ms}$).
This happens because the next-nearest-hopping terms are inversely proportional to the driving frequency.
It is interesting to note that the prediction (\ref{eq:tp}) is close to numerical data  for short times even in this case when the gaps of the effective model are closed, see Fig.~\ref{fig:bands}(c), and the Chern number of the lowest band is not well defined. 
In fact, it is surprising that the anomalous drift even exists in this case, 
as all subbands are now merged into a single band.
We attribute this effect to our choice of the initial state.
When the gaps are closed, it is hard to set the parameters in such a way that the lowest band is completely filled.
The top of this band usually remains empty,
and the particles thus do not ``see'' that the gap is closed.

\begin{figure}
\includegraphics[width=0.475\textwidth]{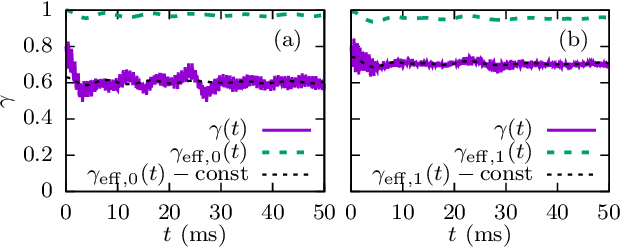}
\caption{Time evolution of the filling factor $\gamma(t)$ for driving frequency $\omega=20$.
Solid purple lines: evolution governed by the time-dependent Hamiltonian $\hat{H}(t)$ 
(cases 2 and 4 from Table \ref{tab:cases}).
Dashed green lines: evolution governed by the effective Hamiltonian $\hat{H}_\mathrm{eff,1}$ or $\hat{H}_\mathrm{eff,0}$ 
(cases 1 and 3).
Dotted black lines: green lines shifted in order to compare them with purple lines.
Shift is chosen so that the two lines approximately overlap.
(a) Initial states and band populations obtained using the
effective Hamiltonian $\hat{H}_\mathrm{eff,0}$, which is without the $J_x^2/\omega$ correction term (cases 3 and 4). 
(b) Hamiltonian $\hat{H}_\mathrm{eff,1}$ which is with the correction term (cases 1 and 2).
\label{fig:gamma}}
\end{figure}

Time evolution of the filling factor $\gamma(t)$ is plotted in Fig.~\ref{fig:gamma}
for four different cases from Table~\ref{tab:cases} -- evolution using 
the effective Hamiltonian without correction $\hat{H}_\mathrm{eff,0}$ [$\gamma_\mathrm{eff,0}(t)$, case 3, dashed green line in Fig.~\ref{fig:gamma}(a)],
the effective Hamiltonian with correction $\hat{H}_\mathrm{eff, 1}$ [$\gamma_\mathrm{eff,1}(t)$, case 1, dashed green line in Fig.~\ref{fig:gamma}(b)], or
the time-dependent Hamiltonian $\hat{H}(t)$ [$\gamma(t)$, cases 2 and 4, solid purple lines].
At the initial moment $\gamma(t_0=0)<1$, 
because the initial state was multiplied by the operator 
$\mathrm{e}^{-i\hat{K}(0)}$.
This introduces a shift between $\gamma(t)$ and $\gamma_\mathrm{eff,1}(t)$.
Apart from the shift, these two curves 
behave similarly, unlike the $\gamma_\mathrm{eff,0}(t)$ curve
that exhibits completely different behavior.
Because of this, we use only $\gamma_\mathrm{eff,1}(t)$ 
to estimate the value of the prediction \eqref{eq:tp}.

We find that the values of $\gamma_{\mathrm{eff,1}}(t)$ for $\omega = 20$ are high: $\geq 0.95$; see Fig.~\ref{fig:gamma}.  For this reason,  up to $50\, \text{ms}$ the center-of-mass position $x(t)$ exhibits roughly linear behavior with some additional oscillations. Interestingly, the anomalous drift $x(t)$ exhibits quadratic behavior on short time scales in all cases from Fig. \ref{fig:drift}.
In Appendix \ref{a:qz}, we explain this feature using the 
time-dependent perturbation theory and Fermi's golden rule.

\section{Interacting case}\label{s:int}

We now investigate the effects of weak repulsive interactions. We work in the high-frequency regime and set $\omega = 20$.
As shown in Sec. \ref{s:bands}, for $U = 0$ the effective Hamiltonian with correction, $\hat{H}_{\mathrm{eff, 1}}$, is in this case equivalent to the Harper-Hofstadter Hamiltonian with flux $\alpha = 1/4$. Moreover, the same approximative form of the full effective model 
accurately reproduces the behavior of the time-dependent Hamiltonian up to $50 \,\text{ms}$ and thus provides a good starting point for the study of weakly interacting particles.
We first consider the anomalous drift of the center of mass of the atomic cloud and then we inspect the expansion dynamics more closely in terms of atomic density distributions in real and momentum space.

\subsection{Anomalous drift and dynamics of band populations}

To simulate the dynamics of many 
 incoherent bosons, we use the classical field method  presented in Sec. \ref{s:incoherent} and propagate Eq.~(\ref{eq:gp}) in time.
We assume that at $t_0 = 0$ atoms are uniformly distributed over the lowest band of $\hat{H}_\mathrm{eff, 1}$. For this reason, the initial state is the same as the one that we use in the noninteracting regime. In this way, the dynamics is initiated by an effective triple quench: at $t_0=0$ the confining potential is turned off,  
atoms are exposed to the force $\mathbf{F}=-F\mathbf{e}_y$, and also the interactions between particles are introduced.
The total number of particles is set to $N=90,000$, 
which amounts to approximately $100$ particles per lattice site in the central region of the atomic cloud. We consider only weak repulsion $U\leq0.05$.

\begin{figure*}[!tb]
\includegraphics[width=0.475\textwidth]{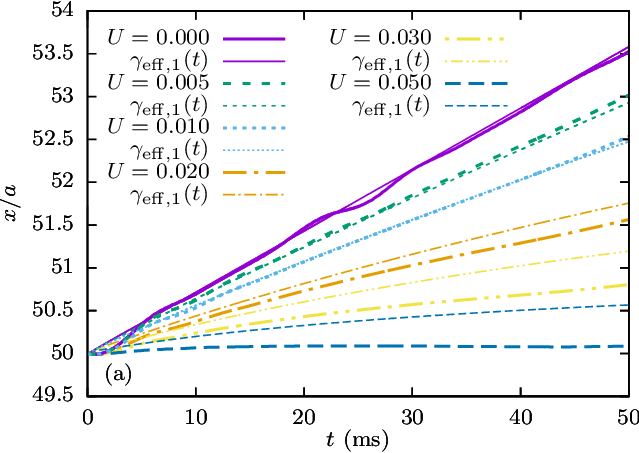}%
\includegraphics[width=0.475\textwidth]{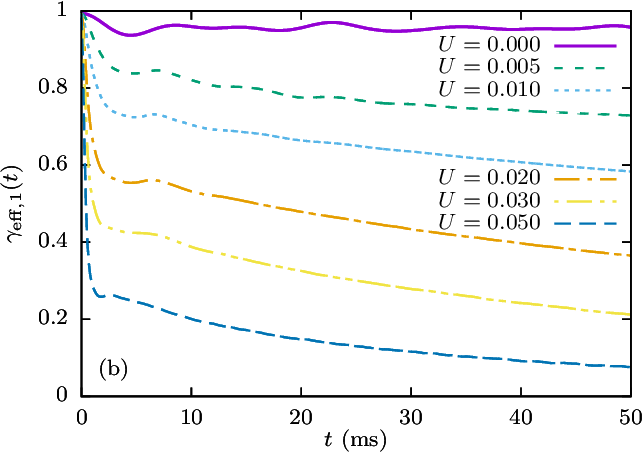}
\caption{Effects of interactions.
(a) Anomalous drift $x(t)$ for several different values of the interaction coefficient $U$.
$U$ is given in units where $J=1$.
Thick lines: numerical simulations using the time-dependent Hamiltonian $\hat{H}(t)$.
Thin lines: theoretical prediction \eqref{eq:tp} from $\gamma_\mathrm{eff,1}(t)$.
(b) Corresponding $\gamma_\mathrm{eff,1}(t)=\eta_1(t)-\eta_2(t)+\eta_3(t)$,
obtained from simulations using the effective Hamiltonian $\hat{H}_\mathrm{eff,1}$. \label{fig:interactions}}
\end{figure*}

The anomalous drift $x(t)$ obtained using the full time-dependent Hamiltonian is shown in Fig.~\ref{fig:interactions}(a) for several different values of the interaction strength $U$. In comparison to the noninteracting regime, we find that the weak repulsive interactions inhibit the response of the center of mass to the external force. In particular, at $t = 50 \, \text{ms}$ the drift is reduced by about $15\%$ for $U = 0.005$ and it is further lowered by an increase in $U$. Finally, at $U = 0.05$, the anomalous drift is barely discernible. Interestingly, for weak $U \in (0.001, 0.01)$ we find that the drift $x(t)$ in the range of $t\in(10,50) \, \text{ms}$ looks ``more linear" as a function of time in comparison to the noninteracting result. 

We now analyze the anomalous drift in terms of the filling factor $\gamma(t)$ and compare the results of Eq.~(\ref{eq:gp}) with the description based on Eq.~(\ref{eq:gp2}).
By solving Eq.~(\ref{eq:gp2}) we obtain the filling factor $\gamma_\mathrm{eff, 1}(t)$ following Eq.~(\ref{eq:eta}) and present our results in Fig. \ref{fig:interactions}(b).
Whenever the results of  Eq.~(\ref{eq:gp}) reasonably agree with the results obtained from Eq.~(\ref{eq:gp2}), we are close to a steady-state regime with only small fluctuations in the total energy, as Eq.~\eqref{eq:gp2} preserves the total energy of the system.
In this regime, during the expansion dynamics the interaction energy is converted into the kinetic energy and atoms are transferred to higher bands of the effective model. Consequently, the filling factor $\gamma_\mathrm{eff,1}(t)$ is reduced. Typically, we find three different stages in the decrease of $\gamma_\text{eff,1}(t)$.

In an early stage,  $t \leq t_1 = 5 \,\text{ms}$, a fast redistribution of particles over the bands of the effective model sets in due to the sudden quench of $U$. The factor $\gamma_\mathrm{eff, 1}(t)$ decays quadratically as a function of time down to $\gamma_\mathrm{eff, 1}(t_1) \approx 0.75$ for $U = 0.01$, and $\gamma_\mathrm{eff, 1}(t_1) \approx 0.25$ for $U = 0.05$. In this process the interaction energy of the system is quickly lowered as described in Appendix \ref{a:en}.
At later times $t> 5\,\text{ms}$, we observe a linear decay of the filling factor $\gamma_\mathrm{eff, 1}(t)$ as a function of time, that finally turns into an exponential decay at even later times ($t>10~\mathrm{ms}$). Similar regimes are observed in other dynamical systems. For example, a decay rate of an initial state suddenly coupled to a bath of additional degrees of freedom exhibits these three stages \cite{Debierre2015}. The initial quadratic decay is often denoted as ``the Zeno regime." For longer propagation times, Fermi's golden rule predicts the linear decay. At even longer time scales, when the repopulation of the initial state is taken into account, the time-dependent perturbation theory yields the exponential regime, known under the name of the Wigner-Weisskopf theory \cite{Debierre2015}.

\begin{figure}[!b]
\includegraphics[width=0.475\textwidth]{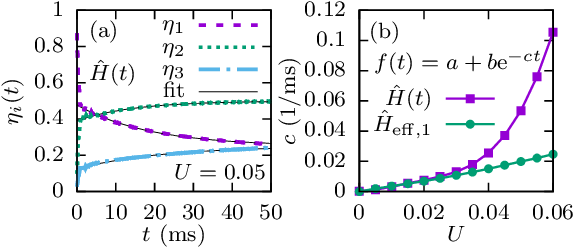}
\caption{(a) Evolution of the band populations $\eta_i(t)$. 
Dashed lines: numerical results obtained using the time-dependent Hamiltonian $\hat{H}(t)$. Solid black lines: exponential fit  using $f(t)=a+b\mathrm{e}^{-ct}$.
The coefficient $a$ was fixed to $a_1=0.25$, $a_2=0.50$ and $a_3=0.25$ for the first, second and third band respectively.
(b) Dependence of the exponential decay coefficients for the lowest band population $\eta_1(t)$ on the interaction strength. 
$U$ is given in units where $J=1$. \label{fig:fit}}
\end{figure}

We now investigate this last regime in more detail. 
For the population of the lowest band $\eta_1(t)$,
an exponential decay function $f(t)=a+b\mathrm{e}^{-ct}$ provides high quality fits for $t\in(10, 50) \, \text{ms}$; see Fig.~\ref{fig:fit}(a) for an example.
Similarly, the populations of two higher bands can also be fitted to exponential functions.
The obtained exponential decay coefficients $c$ for the lowest band population are plotted as a function of the
interaction strength $U$ in Fig.~\ref{fig:fit}(b).
The resulting dependence is approximately quadratic: 
$c(U)=\alpha_0+\alpha_1~U+\alpha_2~U^2$. For small values of $U$, the exponents $c(U)$ obtained for the dynamics governed by $\hat{H}(t)$ and $\hat{H}_\text{eff, 1}$ agree very well and exhibit linear behavior. At stronger interaction strengths $U \geq 0.03$, the approximation of Eq.~(\ref{eq:gp2}) becomes less accurate as it omits the quadratic contribution in $c(U)$ found in the full time evolution.
In addition, the values of the exponents $c$ are affected by the force strength $F$ and driving frequency $\omega$.

As we now understand some basic features of $\gamma_\mathrm{eff, 1}(t)$, we make an explicit comparison between the numerical results for the anomalous drift and the expectation (\ref{eq:tp}).
The dashed lines in Fig.~\ref{fig:interactions}(a) correspond to the theoretical prediction \eqref{eq:tp} calculated from $\gamma_\mathrm{eff,1}(t)$.
For the intermediate interaction strengths $U \leq 0.01$, we find a very good agreement between the two. From this we conclude that the interaction-induced transitions of atoms to higher bands are the main cause of the reduced anomalous drift $x(t)$ as a function of $U$. When the interactions become strong enough ($U\sim0.02$), the numerical results start to deviate from the theoretical prediction \eqref{eq:tp} with $\gamma_\mathrm{eff,1}(t)$. In this regime, Eq.~(\ref{eq:gp2}) does not provide a reliable description of the dynamics, as higher-order corrections need to be taken into account.

\subsection{Real and momentum-space dynamics}\label{s:density}
\begin{figure}[!t]
\includegraphics[width=0.475\textwidth]{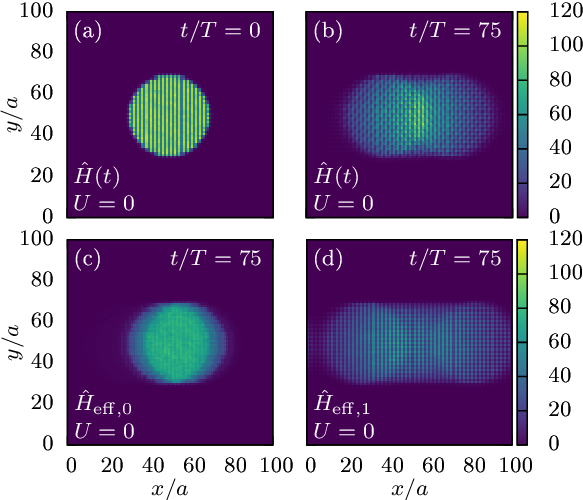}
\caption{
Real-space density distribution, noninteracting case $U=0$.
(a) Initial state.
(b) After $50~\mathrm{ms}$ ($75$ driving periods), evolution using the time-dependent Hamiltonian $\hat{H}(t)$.
(c) Evolution using effective Hamiltonian without correction $\hat{H}_\mathrm{eff,0}$.
(d) Evolution using effective Hamiltonian with correction $\hat{H}_\mathrm{eff,1}$.
\label{fig:density1}}
\end{figure}

So far we have considered the averaged response of the whole atomic cloud. We now inspect the expansion dynamics in a spatially resolved manner.
The real-space probability densities at the initial moment and after $50~\mathrm{ms}$ ($75$ driving periods)
are shown in Figs.~\ref{fig:density1} and \ref{fig:density2}, and the corresponding momentum-space probability densities 
in Appendix \ref{a:kd}.

At the initial moment, the atomic cloud is localized in the center of the lattice. By setting $r_0 = 20$ in the confining potential of Eq.~(\ref{eq:hini}) and populating the lowest-lying states, we fix the cloud radius to $r = 20$, Fig.~\ref{fig:density1}(a). The cloud density is of the order of $100$ atoms per lattice site and a weak density modulation is visible along $x$ direction. After the confining potential is turned off, and the external force in the $-\mathbf{e}_y$ direction is turned on, 
the cloud starts to expand and move in the $+\mathbf{e}_x$ direction.
As shown in the previous subsection, 
the band populations and therefore the anomalous drift
are significantly altered by the interaction strength,
and this is also the case with the expansion dynamics; see Figs.~\ref{fig:density1} and \ref{fig:density2}.

In the noninteracting case, Fig.~\ref{fig:density1}(b), the atomic cloud nearly separates into
two parts moving in opposite directions along $x$ axes (while the center of mass still moves in the $+\mathbf{e}_x$ direction).
By comparing Fig.~\ref{fig:density1}(c) and Fig.~\ref{fig:density1}(d), we conclude that this effect stems from the next-nearest-neighbor hopping along $x$ present in the effective Hamiltonian (\ref{eq:heff}), 
as it does not happen in the effective model without the correction term (\ref{eq:heff0}).
This type of separation was already observed in Ref.~\cite{Dauphin2013},
where the next-nearest-neighbor hopping terms were also present.

\begin{figure}[!t]
\includegraphics[width=0.475\textwidth]{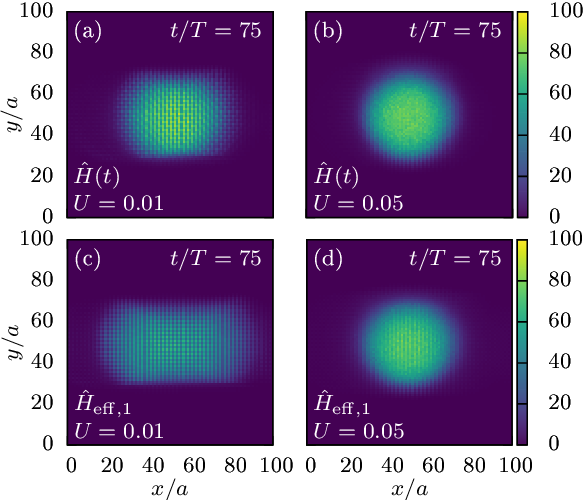}
\caption{
Real-space density distribution after $50~\mathrm{ms}$ ($75$ driving periods), interacting case.
$U$ is given in units where $J=1$.
(a) Evolution using the time-dependent Hamiltonian $\hat{H}(t)$, $U=0.01$.
(b) Same with $U=0.05$.
(c) Evolution using the effective Hamiltonian $\hat{H}_\mathrm{eff,1}$, $U=0.01$.
(d) Same with $U=0.05$.
\label{fig:density2}}
\end{figure}

\begin{figure}[!b]
\includegraphics[width=0.475\textwidth]{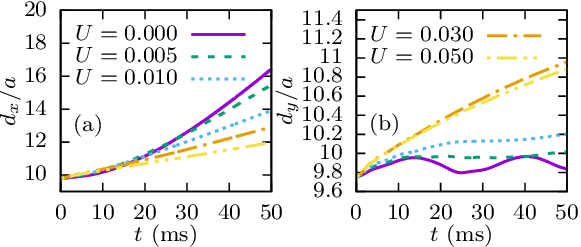}
\caption{Atomic cloud width for different interaction strengths, 
evolution using the time-dependent Hamiltonian $\hat{H}(t)$.
$U$ is given in units where $J=1$.
(a) $d_x=\sqrt{\langle x^2\rangle-\langle x\rangle^2}$.
(b) $d_y=\sqrt{\langle y^2\rangle-\langle y\rangle^2}$.
\label{fig:width}}
\end{figure}

\begin{figure*}[!t]
\includegraphics[width=0.99\textwidth]{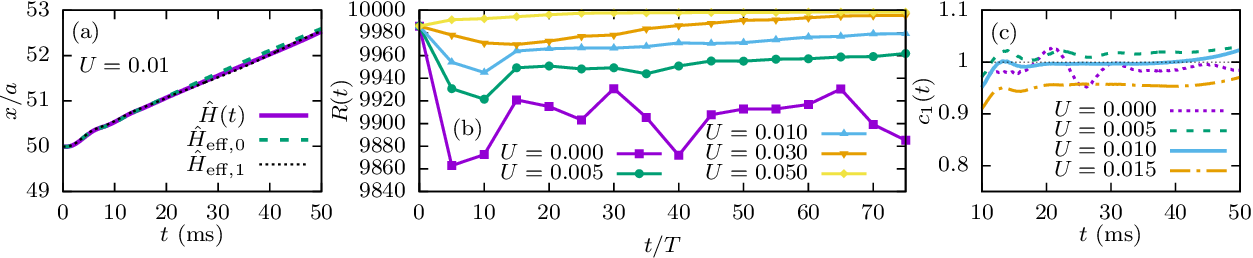}
\caption{(a) Comparison of anomalous drifts obtained from evolution using the time-dependent Hamiltonian $\hat{H}(t)$ (solid purple line),
effective Hamiltonian without correction $\hat{H}_\mathrm{eff,0}$ (dashed green line) and 
effective Hamiltonian with correction $\hat{H}_\mathrm{eff,1}$ (dotted black line). Intermediate interaction strength $U=0.01$.
$U$ is given in units where $J=1$.
(b) Time evolution of the inverse participation ratio in momentum space for several different values of $U$. 
Evolution is performed using the time-dependent Hamiltonian $\hat{H}(t)$. 
When the interactions are strong enough, 
$\mathrm{IPR}$ approaches the maximal possible value ($10,000$ in this case),
which is equal to the total number of states and corresponds to the completely delocalized state.
$U$ is given in units where $J=1$.
(c) Chern number of the lowest band obtained for different interaction strengths as the ratio of the theoretical prediction for the anomalous drift and numerical results:
$c_1(t)=\left(\frac{2Fa^2}{\pi\hbar}\int_0^t \gamma_\mathrm{eff,1}(t')dt'\right)/\left(x(t)-x(t_0)\right)$.
\label{fig:ipr-chern}}
\end{figure*}
When the interactions between particles are included, this separation is not so prominent 
[Fig.~\ref{fig:density2}(a), $U=0.01$], and it almost completely disappears when the interactions
are strong enough [Fig.~\ref{fig:density2}(b), $U=0.05$]. 
This is also the case when the evolution is governed by the effective Hamiltonian $\hat{H}_\mathrm{eff,1}$; see Figs.~\ref{fig:density2}(c) and \ref{fig:density2}(d).
Atomic cloud widths $d_x=\sqrt{\langle x^2\rangle-\langle x\rangle^2}$ during the expansion are plotted in Fig.~\ref{fig:width}.
We observe a slow expansion of the cloud in $y$ direction, Fig.~\ref{fig:width}(b),  and much faster expansion along $x$ direction, Fig.~\ref{fig:width}(a), which comes about as a consequence of the cloud separation. On top of this, we observe that the interactions enhance expansion along $y$. Surprisingly, the opposite is true for the dynamics along $x$. This counterintuitive effect is often labeled as self-trapping and its basic realization is known for the double-well potential \cite{Milburn1997, Raghavan1999}. In brief, strong repulsive interactions can preserve the density imbalance between the two wells, as the system can not release an excess of the interaction energy. In our case, the situation is slightly more involved as the cloud splitting is inherent (induced by the corrections of the ideal effective Hamiltonian). Apart from this, due to the driving the total energy is not conserved. However, our numerical results indicate that the interaction energy is slowly released in the second expansion stage, Fig.~\ref{fig:energy}. Effectively, in this way the interactions cancel out the contribution of the next-nearest-neighbor hopping and favor the measurement of the properties of the model (\ref{eq:heff0}). In Fig.~\ref{fig:ipr-chern}(a) we show that deviations between different approximations based on $\hat{H}(t)$, $\hat{H}_\text{eff, 1}$, and $\hat{H}_\text{eff, 0}$ in the anomalous drift $x(t)$ nearly vanish at $U = 0.01$.

Another desirable effect might be that the interactions make the momentum-space probability density
more homogeneous, see Appendix \ref{a:kd}, 
so that the real-space probability density becomes more localized.
We can quantify momentum-space homogeneity using the inverse participation ratio 
$R(t)=\frac{1}{\sum_i P_i^2(t)}$, where $P_i(t)=\lvert \psi_i(t)\lvert^2$
is the probability that the state $\psi_i$ is occupied at time $t$.
Minimal value of the inverse participation ratio ($\mathrm{IPR}$) is $1$ and it corresponds to a completely localized state, 
while the maximal value is equal to the total number of states (in our case $10,000$)
and corresponds to the completely delocalized state, 
where the particles have the same probability of being at any quasimomentum $\mathbf{k}$.
As stated before,
the first Brillouin zone of the lowest band has to be as homogeneously populated as possible
in order to properly measure the lowest band Chern number.
From Fig.~\ref{fig:ipr-chern}(b), we see that $\mathrm{IPR}$ increases in time when the interaction coefficient $U$
is large enough, so we can conclude that the interactions are actually beneficial for measuring the Chern number,
as they can ``smooth-out'' the momentum-space probability density. In Fig.~\ref{fig:ipr-chern}(c) we give estimates for the Chern number that can be extracted from our numerical data for different values of $U$. We find the best estimate $c_1 \sim 0.99$ for the intermediate interaction strength $U \sim 0.01$.

\subsection{Staggered detuning}\label{s:detuning}

\begin{figure}[!t]
\includegraphics[width=0.475\textwidth]{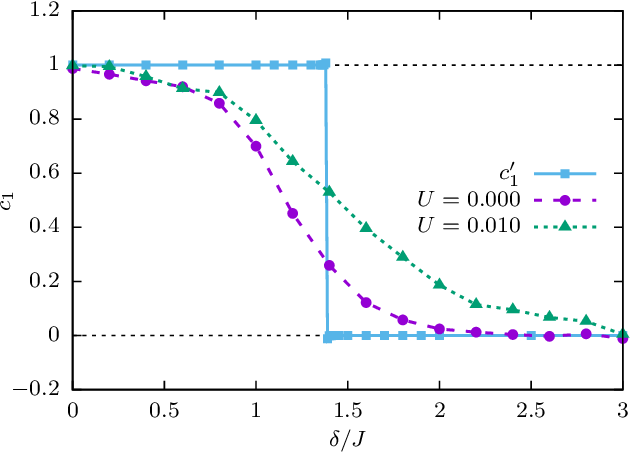}
\caption{Lowest band Chern numbers extracted from numerical data for several different values of detuning $\delta$. Purple circles: noninteracting case, $U=0$. Green triangles: $U=0.01$.
Blue squares: Theoretical values of the lowest band Chern number $c'_1$. A topological phase transition is visible at $\delta_c\approx 1.38$. The lines between points are only a guide to the eye.
\label{fig:detuning}}
\end{figure}

Here we briefly consider the effects of staggered detuning that was introduced in the experimental study \cite{Aidelsburger2015} during the loading and band mapping sequences. This detuning can be described by an additional term
\begin{equation}\label{eq:detuning}
\frac{\delta}{2}\sum_{l,m}\left[(-1)^l+(-1)^m\right]\hat{n}_{l,m}
\end{equation}
in the Hamiltonians $\hat{H}(t)$ and $\hat{H}_\mathrm{eff,1}$. We will ignore the higher-order [at most $\mathcal{O}\left(\frac{1}{\omega^2}\right)$] corrections that this term introduces to the effective Hamiltonian.
Staggered detuning does not break the symmetry of the effective Hamiltonian $\hat{H}_\mathrm{eff,1}$, but if $\delta$ is large enough, it can cause a topological phase transition and make all bands topologically trivial. 
By numerically calculating the Berry curvature and Chern numbers $c'_i$, we find that this transition occurs at $\delta_c\approx 1.38\,J$; see Fig.~\ref{fig:detuning}. This value is lower than the one for the ordinary Harper-Hofstadter Hamiltonian for $\alpha=1/4$, which is $\delta_c=2\,J$ \cite{Aidelsburger2015}, due to the different hopping amplitudes $J'_x$ and $J'_y$, 
and due to the additional $J_x^2/\omega$ correction that we consider.

We now investigate how this topological transition can be probed through the dynamical protocol used in the experiment.
We again numerically calculate the anomalous drift and the evolution of the filling factor, 
but now with staggered detuning \eqref{eq:detuning} included in the Hamiltonian $\hat{H}_\mathrm{initial}$ \eqref{eq:hini} used to obtain the initial state, in the equations of motion \eqref{eq:gp} and \eqref{eq:gp2}, and in the definitions of the band populations $\eta_i(t)$ \eqref{eq:eta}. 
Using these results, we repeat the procedure for the extraction of the lowest band Chern number from numerical data that was carried out in the previous section. The Chern number obtained by comparing the anomalous drift to the prediction calculated from the filling factor is then averaged over the time interval $t\in(20, 40) \, \text{ms}$. This interval was chosen in order to avoid the initial quadratic regime and the finite-size effects at later times.
The resulting lowest band Chern numbers for several different values of detuning $\delta$ in both the noninteracting case and the case of intermediate interaction strength $U=0.01$ are presented in Fig~\ref{fig:detuning}.

We can see that the calculated value of the Chern number decreases from $c_1=1$ to $c_1=0$
with increasing detuning $\delta$. 
The obtained value of the Chern number is lower than $1$ even before the phase transition occurs. This is due to our choice of the initial state, which is not perfectly homogeneous in momentum space. Close to the phase transition, both the energy bands and the Berry curvature have pronounced peaks at the same regions of the first Brillouin zone, and these regions are initially less populated. Because of this, the Berry curvature at these regions contributes less to the anomalous drift, which lowers the measured Chern number.
This effect is somewhat reduced by the interactions, as they smooth out
the momentum-space probability density, and might also cancel out the detuning term. 
Similar interplay of interactions and staggering was observed in the fermionic Hofstadter-Hubbard model \cite{Cocks2012}.
The obtained results are in line with 
experimental measurements \cite{Aidelsburger2015}.

\section{Conclusions}\label{s:conclusions}

Motivated by the recent experimental results reporting the Chern numbers of topological bands in cold-atom setups, we studied numerically bosonic transport in a driven optical lattice. 
The considered driving scheme and the range of microscopic parameters were chosen to be close to those in a recent experimental study \cite{Aidelsburger2015}.
The driving frequency was set to be high enough in order to avoid strong energy absorption for the relevant time scales.
Additionally, the system was restricted to a two-dimensional lattice, even though the actual experimental setup had continuous transverse degrees of freedom. 
This restriction stabilizes the system \cite{Bilitewski2015,Choudhury2015,Lellouch2017} and leads to lower heating rates than those in the experiment. It corresponds to the case of strongly confined third dimension.

We investigated bosonic dynamics for the full time-dependent Hamiltonian, the effective Floquet Hamiltonian, and included the effects of weak repulsive interactions between atoms using the mean-field approximation.
In the noninteracting case, we found that the effective Hamiltonian and its band structure depend on the frequency of the drive $\omega$ through an additional $J_x^2 /\omega$ correction term. 
The initial state was set as a mixture of incoherent bosons homogeneously populating the lowest band, 
but a possible direction of future research could be to 
simulate the full loading sequence of an initial Bose-Einstein condensate
and to try to obtain the incoherent state through driving, 
as it was done in the experiment.

The main focus of this work is on the effects of weak interactions.
For a weak atomic repulsion, atomic transitions to higher effective bands obtained in our simulations mainly occur due to a release of the initial interaction energy during the atomic-cloud expansion. Although the effect is undesirable, it can be properly taken into account in the extraction of the Chern number.
At larger interaction strengths, the transitions are more pronounced as the system absorbs energy from the drive. 
In this regime the good agreement between the full and effective description is lost and the measurement should become more complicated.
In addition to causing redistribution of atoms over bands, our results show that weak interactions can also be beneficial in measuring the Chern number. Their desirable effect comes about due to smoothening the atomic distribution over the topological band and due to canceling out the contribution of some less relevant terms to the bosonic dynamics.

\acknowledgments
This work was supported by the Ministry of Education, Science, and
Technological Development of the Republic of Serbia under Projects
No.~ON171017, BKMH, and TOP-FOP, the Croatian Science Foundation under
Grant No.~IP-2016-06-5885 SynthMagIA and the TOP-FOP project, and
by DAAD (German Academic and Exchange Service) under the BKMH
project.
Numerical simulations were performed on the PARADOX supercomputing
facility at the Scientific Computing Laboratory of the Institute
of Physics Belgrade.
This research was funded by the Deutsche Forschungsgemeinschaft
(DFG, German Research Foundation) via Research Unit FOR 2414 under
project number 277974659.
This work was also supported by the Deutsche
Forschungsgemeinschaft (DFG) via the high-performance computing
center LOEWE-CSC.

\appendix
\begin{widetext}

\section{The effective model}
\label{a:heff}


After a unitary transformation into the rotating frame $\tilde{\psi}=\mathrm{e}^{-i\hat{W}t}\psi$,
where $\tilde{\psi}$ and $\psi$ are the old and the new wave functions,
and $\hat{W}$ is the staggered potential,
the new time-dependent Hamiltonian that describes the experimental setup
is given by \cite{Aidelsburger2015}
\begin{equation}\label{eq:htd}
 \hat{H}(t)=J_y\sum_{l,m} \left(\hat{a}^\dagger_{l,m+1}\hat{a}_{l,m}+\hat{a}^\dagger_{l,m-1}\hat{a}_{l,m}\right)
 +\hat{V}^{(+1)}\mathrm{e}^{i\omega t}+\hat{V}^{(-1)}\mathrm{e}^{-i\omega t}+\frac{U}{2}\sum_{l,m}\hat{n}_{l,m}\left(\hat{n}_{l,m}-1\right),
\end{equation}
where
\begin{equation}\label{eq:vp}
 \hat{V}^{(+1)}=\kappa/2\sum_{l,m} \hat{n}_{l,m} g(l,m)-J_x\sum_{l_{\mathrm{odd}},m} \left(\hat{a}^\dagger_{l+1,m}\hat{a}_{l,m}+\hat{a}^\dagger_{l-1,m}\hat{a}_{l,m}\right)
 \end{equation}
\begin{equation}\label{eq:vm}
 \hat{V}^{(-1)}=\kappa/2\sum_{l,m} \hat{n}_{l,m} g^*(l,m)-J_x\sum_{l_{\mathrm{even}},m} \left(\hat{a}^\dagger_{l+1,m}\hat{a}_{l,m}+\hat{a}^\dagger_{l-1,m}\hat{a}_{l,m}\right)
\end{equation}
\begin{equation}
 g(l,m)=\cos(l\pi/2-\pi/4)\mathrm{e}^{i(\phi_0-m\pi/2)}+\cos(l\pi/2+\pi/4)\mathrm{e}^{i(m\pi/2-\phi_0-\pi/2)}.
\end{equation}
The kick operator is given by
\begin{equation}\label{eq:kick}
 \hat{K}(t)=\frac{1}{i\omega}\left(\hat{V}^{(+1)}\mathrm{e}^{i\omega t}-\hat{V}^{(-1)}\mathrm{e}^{-i\omega t}\right)+\mathcal{O}\left(\frac{1}{\omega^2}\right),
\end{equation}
and the effective Hamiltonian by
\begin{align}\label{eq:commutators}
\nonumber
 \hat{H}_{\mathrm{eff}}=&\underbrace{\hat{H}_{0}}_{\hat{H}_{\mathrm{eff}}^{(0)}}
 +\underbrace{\frac{1}{\omega}\left[\hat{V}^{(+1)},\hat{V}^{(-1)}\right]}_{\hat{H}_{\mathrm{eff}}^{(1)}}\\
 &+\underbrace{\frac{1}{2\omega^2}\left(\left[\left[\hat{V}^{(+1)},\hat{H}_{0}\right],\hat{V}^{(-1)}\right]+
 \left[\left[\hat{V}^{(-1)},\hat{H}_{0}\right],\hat{V}^{(+1)}\right]\right)}_{\hat{H}_{\mathrm{eff}}^{(2)}}
 +\mathcal{O}\left(\frac{1}{\omega^3}\right).
\end{align}

If we assume that the driving frequency is high and interactions are weak,
the interaction term and almost all $\mathcal{O}\left(\frac{1}{\omega^2}\right)$ terms can be neglected.
After substituting Eqs. \eqref{eq:htd}, \eqref{eq:vp} and \eqref{eq:vm} into Eq. \eqref{eq:commutators} we obtain:
\begin{align}
 \hat{H}_{\mathrm{eff}}^{(0)}=&-J_y\sum_{l,m} 
 \left(\hat{a}^\dagger_{l,m+1}\hat{a}_{l,m}
 +\hat{a}^\dagger_{l,m-1}\hat{a}_{l,m}\right)\\
 \nonumber
 \hat{H}_{\mathrm{eff}}^{(1)}=&\frac{1}{\omega}\bigg[
 \frac{\kappa}{2}\sum_{l,m} \hat{a}^\dagger_{l,m}\hat{a}_{l,m} ~g(l,m) 
 -J_x\sum_{l_{\mathrm{odd}},m} \left(\hat{a}^\dagger_{l+1,m}\hat{a}_{l,m}
 +\hat{a}^\dagger_{l-1,m}\hat{a}_{l,m}\right),\\
 &\frac{\kappa}{2}\sum_{l,m} \hat{a}^\dagger_{l,m}\hat{a}_{l,m} ~g^*(l,m) 
 -J_x\sum_{l_{\mathrm{even}},m} \left(\hat{a}^\dagger_{l+1,m}\hat{a}_{l,m}
 +\hat{a}^\dagger_{l-1,m}\hat{a}_{l,m}\right)\bigg]\\
 =&\hspace{0.1cm} \hat{H}_1+\hat{H}_2+\hat{H}_3+\hat{H}_4.\nonumber
 \end{align}
 We will now separately calculate each term: 
 \begin{align}\label{eq:I}
 \nonumber
 \hat{H}_1=&-\frac{J_x\kappa}{2\omega}\sum_{l_{\mathrm{odd}},m,l',m'}g^*(l',m')
 \Big[\hat{a}^\dagger_{l+1,m}\hat{a}_{l,m}+\hat{a}^\dagger_{l-1,m}\hat{a}_{l,m},
 \hat{a}^\dagger_{l',m'}\hat{a}_{l',m'}\Big]\\
 =&-\frac{J_x\kappa}{2\omega}\sum_{l_{\mathrm{odd}},m}
 \Big[\big(g^*(l,m)-g^*(l+1,m)\big)
 \hat{a}^\dagger_{l+1,m}\hat{a}_{l,m}
 +\big(g^*(l,m)-g^*(l-1,m)\big)
 \hat{a}^\dagger_{l-1,m}\hat{a}_{l,m}\Big]
 \end{align}
 \begin{align}\label{eq:II}
  \nonumber
 \hat{H}_2=&-\frac{J_x\kappa}{2\omega}\sum_{l_{\mathrm{even}},m,l',m'}g(l',m')
 \Big[\hat{a}^\dagger_{l',m'}\hat{a}_{l',m'},
 \hat{a}^\dagger_{l+1,m}\hat{a}_{l,m}+\hat{a}^\dagger_{l-1,m}\hat{a}_{l,m}\Big]\\
 =&\frac{J_x\kappa}{2\omega}\sum_{l_{\mathrm{even}},m}
 \Big[\big(g(l,m)-g(l+1,m)\big)
 \hat{a}^\dagger_{l+1,m}\hat{a}_{l,m}
 +\big(g(l,m)-g(l-1,m)\big)
 \hat{a}^\dagger_{l-1,m}\hat{a}_{l,m}\Big]
 \end{align}
 \begin{align}\label{eq:III}
   \nonumber
 \hat{H}_3=&\frac{J^2_x}{\omega}\sum_{l_{\mathrm{odd}},m,l'_{\mathrm{even}},m'}
 \Big[\hat{a}^\dagger_{l+1,m}\hat{a}_{l,m}+\hat{a}^\dagger_{l-1,m}\hat{a}_{l,m},
 \hat{a}^\dagger_{l'+1,m'}\hat{a}_{l',m'}+\hat{a}^\dagger_{l'-1,m'}\hat{a}_{l',m'}\Big]\\
  =&\frac{J^2_x}{\omega}\sum_{l_{\mathrm{odd}},m}  \nonumber
 \Big(2\hat{a}^\dagger_{l+1,m}\hat{a}_{l+1,m}
 +\hat{a}^\dagger_{l+3,m}\hat{a}_{l+1,m}
 +\hat{a}^\dagger_{l-1,m}\hat{a}_{l+1,m}\\  \nonumber
 &\hspace{1.6cm}-2\hat{a}^\dagger_{l,m}\hat{a}_{l,m}
 -\hat{a}^\dagger_{l+2,m}\hat{a}_{l,m}
 -\hat{a}^\dagger_{l-2,m}\hat{a}_{l,m}\Big)\\
 =&\frac{J^2_x}{\omega}\sum_{l,m}(-1)^l
 \Big(2\hat{a}^\dagger_{l,m}\hat{a}_{l,m}
 +\hat{a}^\dagger_{l+2,m}\hat{a}_{l,m}
 +\hat{a}^\dagger_{l-2,m}\hat{a}_{l,m}\Big)
 \end{align}
 \begin{align}
 \hat{H}_4=&\frac{\kappa^2}{4\omega}\sum_{l,m,l',m'}g(l,m)g^*(l',m')
 \Big[\hat{a}^\dagger_{l,m}\hat{a}_{l,m},\hat{a}^\dagger_{l',m'}\hat{a}_{l',m'}\Big]=0.
 \end{align}
 Using trigonometric identities and
 \begin{align}
 g(l,m)-g(l\pm1,m)=&\pm\sqrt{2}\Big(\sin((2l\pm1-1)\pi/4)\mathrm{e}^{i(\pi/4-m\pi/2)}\\
 &+\sin((2l\pm1+1)\pi/4)\mathrm{e}^{i(m\pi/2-3\pi/4)}\Big),
\end{align}
we can rewrite the sum of terms \eqref{eq:I} and \eqref{eq:II}
in a more convenient form
 \begin{align}
 \hat{H}_1+\hat{H}_2=&\frac{J_x\kappa}{\sqrt{2}\omega}\sum_{l,m}
 \Big(\mathrm{e}^{i\big((m-l)\pi/2-\pi/4\big)}
 \hat{a}^\dagger_{l,m}\hat{a}_{l-1,m}
 +\mathrm{e}^{-i\big((m-l-1)\pi/2-\pi/4\big)}
 \hat{a}^\dagger_{l,m}\hat{a}_{l+1,m}\Big).
 \end{align}
 The only $\mathcal{O}\left(\frac{1}{\omega^2}\right)$ ($\hat{H}_{\mathrm{eff}}^{(2)}$) term 
 that cannot be neglected in the parameter range that we use is \cite{Aidelsburger2015}
 \begin{align}
 \frac{J_y}{2}\frac{\kappa^2}{\omega^2}\sum_{l,m} 
 \left(\hat{a}^\dagger_{l,m+1}\hat{a}_{l,m}
 +\hat{a}^\dagger_{l,m-1}\hat{a}_{l,m}\right).
 \end{align}
 
 Finally, the effective Hamiltonian becomes
 \begin{align}
 \nonumber
 \hat{H}_{\mathrm{eff,1}}=&\frac{J_x\kappa}{\sqrt{2}\omega}\sum_{l,m}
 \Big(\mathrm{e}^{i\big((m-l-1)\pi/2-\pi/4\big)}
 \hat{a}^\dagger_{l+1,m}\hat{a}_{l,m}
 +\mathrm{e}^{-i\big((m-l)\pi/2-\pi/4\big)}
 \hat{a}^\dagger_{l-1,m}\hat{a}_{l,m}\Big)\\
 &-J_y\Big(1-\frac{1}{2}\frac{\kappa^2}{\omega^2}\Big)\sum_{l,m} 
 \left(\hat{a}^\dagger_{l,m+1}\hat{a}_{l,m}
 +\hat{a}^\dagger_{l,m-1}\hat{a}_{l,m}\right)\\ \label{eq:corrction}
 &+\frac{J^2_x}{\omega}\sum_{l,m}(-1)^l
 \Big(2\hat{a}^\dagger_{l,m}\hat{a}_{l,m}
 +\hat{a}^\dagger_{l+2,m}\hat{a}_{l,m}
 +\hat{a}^\dagger_{l-2,m}\hat{a}_{l,m}\Big)
 \end{align}
 with the renormalized nearest-neighbor hopping amplitudes $J'_x=\frac{J_x\kappa}{\sqrt{2}\omega}=J_y$
 and $J'_y=J_y\Big(1-\frac{1}{2}\frac{\kappa^2}{\omega^2}\Big)$,
 and a next-nearest-neighbor along $\mathbf{e}_x$ hopping term 
 proportional to $\frac{J^2_x}{\omega}$ in \eqref{eq:corrction}.

\section{Effective Hamiltonian in momentum-space}\label{a:hk}

\begin{figure}[!b]
\includegraphics[width=0.99\textwidth]{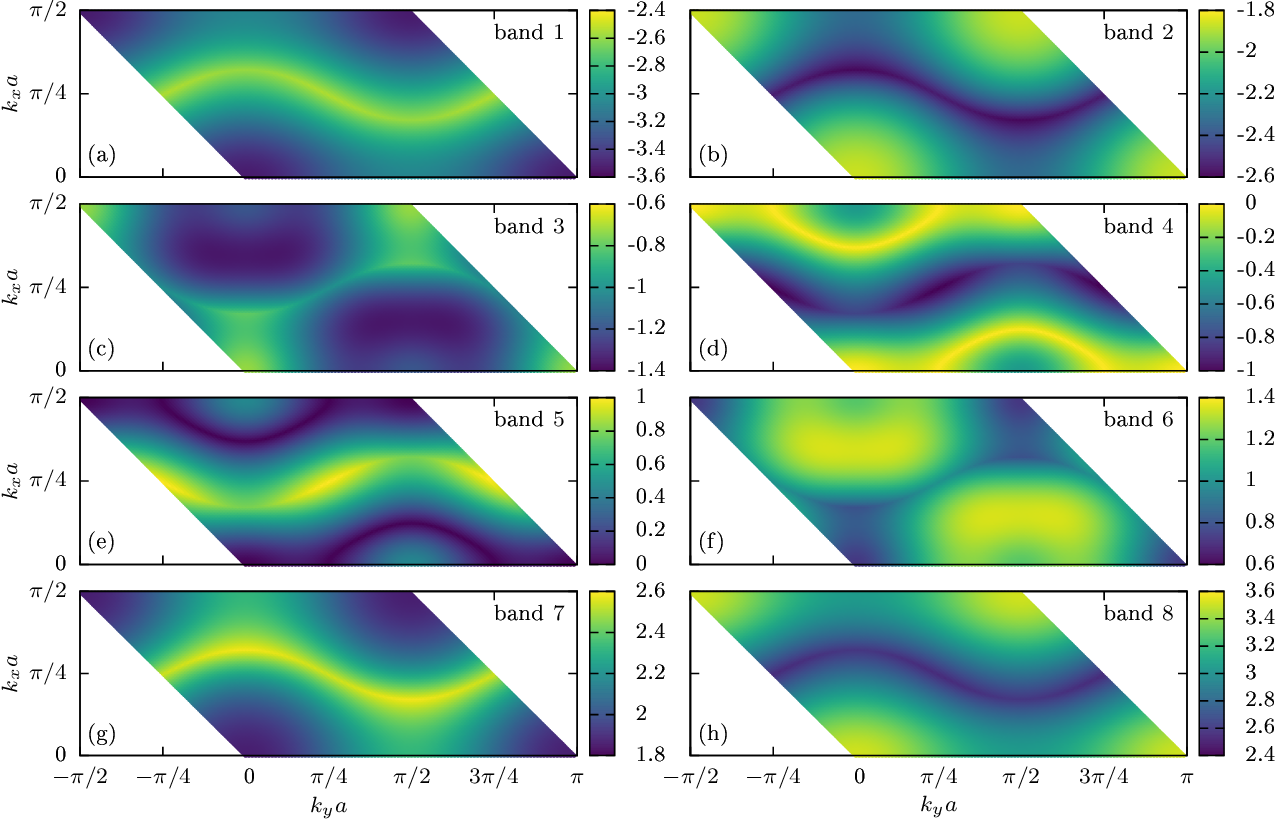}
\caption{Eight energy subbands of $\hat{\mathcal{H}}_{\mathrm{eff,1}}(\mathbf{k})$ for the driving frequency $\omega=20$.
Subbands $1$ and $2$ form the lowest band with Chern number $c_1=1$, 
subbands $3$, $4$, $5$, and $6$ form the middle band with $c_2=-2$,
and subbands $7$ and $8$ form the highest band with $c_3=1$.
\label{fig:bands-xy}}
\end{figure}

If we choose the unit cell as in Fig.~\ref{fig:lattice}(a) 
(lattice sites ${\tt A}=(1,0)$, ${\tt B}=(2,0)$, ${\tt C}=(3,0)$ and ${\tt D}=(4,0)$),
the momentum-space representation of the effective Hamiltonian without correction 
$\hat{H}_\mathrm{eff,0}$ \eqref{eq:heff0} is given by a $4\times4$ matrix
\begin{align}
\tiny 
\hat{\mathcal{H}}_{\mathrm{eff,0}}(\mathbf{k})=
  \begin{psmallmatrix}\label{eq:hk0}
    0 & J'_x\mathrm{e}^{-i\frac{3\pi}{4}}-J'_y\mathrm{e}^{-i\mathbf{k}\mathbf{R}_2} & 0 & J'_x\mathrm{e}^{-i\frac{3\pi}{4}-i\mathbf{k}\mathbf{R}_1}-J'_y\mathrm{e}^{i\mathbf{k}(\mathbf{R}_2-\mathbf{R}_1)}  \\
    J'_x\mathrm{e}^{i\frac{3\pi}{4}}-J'_y\mathrm{e}^{i\mathbf{k}\mathbf{R}_2} & 0 & J'_x\mathrm{e}^{-i\frac{\pi}{4}}-J'_y\mathrm{e}^{-i\mathbf{k}\mathbf{R}_2} & 0  \\
    0 & J'_x\mathrm{e}^{i\frac{\pi}{4}}-J'_y\mathrm{e}^{i\mathbf{k}\mathbf{R}_2} & 0 & J'_x\mathrm{e}^{i\frac{\pi}{4}}-J'_y\mathrm{e}^{-i\mathbf{k}\mathbf{R}_2}  \\
    J'_x\mathrm{e}^{i\frac{3\pi}{4}+i\mathbf{k}\mathbf{R}_1}-J'_y\mathrm{e}^{i\mathbf{k}(\mathbf{R}_1-\mathbf{R}_2)} & 0 & J'_x\mathrm{e}^{-i\frac{\pi}{4}}-J'_y\mathrm{e}^{i\mathbf{k}\mathbf{R}_2} & 0
  \end{psmallmatrix},
\end{align}
where $\mathbf{R}_1$ and $\mathbf{R}_2$ are the lattice vectors $\mathbf{R}_1=(4,0)$ and $\mathbf{R}_2=(1,1)$,
and $\mathbf{k}$ is in the first Brillouin zone, which is given by the reciprocal lattice vectors 
$\mathbf{b}_1=\frac{\pi}{2}(1,-1)$ and $\mathbf{b}_2=2\pi(0,1)$.

When the $\frac{J^2_x}{\omega}$ correction is included in the effective Hamiltonian, 
$\hat{H}_\mathrm{eff,1}$ \eqref{eq:heff},
the unit cell is doubled, see Fig.~\ref{fig:lattice}(b), 
and the first Brillouin zone is therefore halved. 
If we now choose the lattice sites 
${\tt a}=(1,0)$, ${\tt B}=(2,0)$, ${\tt c}=(3,0)$, ${\tt D}=(4,0)$, 
${\tt A}=(2,1)$, ${\tt b}=(3,1)$, ${\tt C}=(4,1)$ and ${\tt d}=(5,1)$ for the unit cell,
the momentum-space representation of the effective Hamiltonian will be an $8\times8$ matrix
\begin{align}\label{eq:hk}
\tiny
\hat{\mathcal{H}}_{\mathrm{eff,1}}(\mathbf{k})=
  \begin{psmallmatrix}
    -\frac{2J_x^2}{\omega} & J_x'\mathrm{e}^{-i\frac{3\pi}{4}} & -\frac{J_x^2}{\omega}(1+\mathrm{e}^{i\mathbf{k}\mathbf{R}_1}) & J_x'\mathrm{e}^{-i(\frac{3\pi}{4}-\mathbf{k}\mathbf{R}_1)} & 0 & -J_y'\mathrm{e}^{i\mathbf{k}\mathbf{R}_2} & 0 & -J_y'\mathrm{e}^{i\mathbf{k}\mathbf{R}_1} \\
    J_x'\mathrm{e}^{i\frac{3\pi}{4}} & \frac{2J_x^2}{\omega} & J_x'\mathrm{e}^{-i\frac{\pi}{4}} & \frac{J_x^2}{\omega}(1+\mathrm{e}^{i\mathbf{k}\mathbf{R}_1}) & -J_y' & 0 & -J_y'\mathrm{e}^{i\mathbf{k}\mathbf{R}_2} & 0 \\
    -\frac{J_x^2}{\omega}(1+\mathrm{e}^{-i\mathbf{k}\mathbf{R}_1}) & J_x'\mathrm{e}^{i\frac{\pi}{4}} & -\frac{2J_x^2}{\omega} & J_x'\mathrm{e}^{i\frac{\pi}{4}} & 0 & -J_y' & 0 & -J_y'\mathrm{e}^{i\mathbf{k}\mathbf{R}_2} \\
    J_x'\mathrm{e}^{i(\frac{3\pi}{4}-\mathbf{k}\mathbf{R}_1)} & \frac{J_x^2}{\omega}(1+\mathrm{e}^{-i\mathbf{k}\mathbf{R}_1}) & J_x'\mathrm{e}^{-i\frac{\pi}{4}} & \frac{2J_x^2}{\omega} & -J_y'\mathrm{e}^{-i\mathbf{k}(\mathbf{R}_1-\mathbf{R}_2)} & 0 & -J_y' & 0 \\
    0 & 0 & 0 & -J_y'\mathrm{e}^{i\mathbf{k}(\mathbf{R}_1-\mathbf{R}_2)} & \frac{2J_x^2}{\omega} & J_x'\mathrm{e}^{-i\frac{3\pi}{4}} & \frac{J_x^2}{\omega}(1+\mathrm{e}^{i\mathbf{k}\mathbf{R}_1}) & J_x'\mathrm{e}^{-i(\frac{3\pi}{4}-\mathbf{k}\mathbf{R}_1)} \\
    -J_y'\mathrm{e}^{-i\mathbf{k}\mathbf{R}_2} & 0 & -J_y' & 0 & J_x'\mathrm{e}^{i\frac{3\pi}{4}} & -\frac{2J_x^2}{\omega} & J_x'\mathrm{e}^{-i\frac{\pi}{4}} & -\frac{J_x^2}{\omega}(1+\mathrm{e}^{i\mathbf{k}\mathbf{R}_1}) \\
    0 &  -J_y'\mathrm{e}^{-i\mathbf{k}\mathbf{R}_2} & 0 & -J_y' & \frac{J_x^2}{\omega}(1+\mathrm{e}^{-i\mathbf{k}\mathbf{R}_1}) & J_x'\mathrm{e}^{i\frac{\pi}{4}} & \frac{2J_x^2}{\omega} & J_x'\mathrm{e}^{i\frac{\pi}{4}} \\
    -J_y'\mathrm{e}^{-i\mathbf{k}\mathbf{R}_1} & 0 &  -J_y'\mathrm{e}^{-i\mathbf{k}\mathbf{R}_2} & 0 & J_x'\mathrm{e}^{i(\frac{3\pi}{4}-\mathbf{k}\mathbf{R}_1)} & -\frac{J_x^2}{\omega}(1+\mathrm{e}^{-i\mathbf{k}\mathbf{R}_1}) & J_x'\mathrm{e}^{-i\frac{\pi}{4}} & -\frac{2J_x^2}{\omega} 
  \end{psmallmatrix},
\end{align}
with the lattice vectors $\mathbf{R}_1=(4,0)$ and $\mathbf{R}_2=(2,2)$.
The reciprocal lattice vectors are then $\mathbf{b}_1=\frac{\pi}{2}(1,-1)$ and $\mathbf{b}_2=\pi(0,1)$.

The energy bands of $\hat{\mathcal{H}}_{\mathrm{eff,1}}(\mathbf{k})$ 
are shown in Figs. \ref{fig:bands} and \ref{fig:bands-xy}.

\section{Description of incoherent bosons}\label{a:ib}

In a typical condensed-matter system constituent particles are electrons. 
Due to their fermionic statistics, at low enough temperatures, and with Fermi energy above the lowest band, that band of the topological model is
uniformly occupied, and consequently the transverse Hall conductivity can be expressed in terms of the Chern number (\ref{eq:cn}) \cite{Thouless1982}. 
In contrast, weakly interacting bosons in equilibrium form a Bose-Einstein condensate in the band minima and only probe the local Berry curvature \cite{Price2012}. 

Yet in the experiment \cite{Aidelsburger2015} the Chern number was successfully measured using bosonic atoms of ${}^{87}\mathrm{Rb}$. This was possible because in the process of ramping up the drive (\ref{eq:drive}), the initial Bose-Einstein condensate was transferred into an incoherent bosonic mixture. Conveniently, it turned out that the bosonic distribution  over the states of the lowest band of the effective Floquet Hamiltonian was nearly uniform. 
Motivated by the experimental procedure, we model the initial bosonic state by a statistical matrix 
\begin{equation}\label{eq:rhoini}
 \rho(t = 0) = \prod_{k=1}^{N_m}  |k, N_p\rangle \langle k, N_p|
\end{equation}
where the states $|k\rangle = a_k^{\dagger} |0\rangle$ approximately correspond to the lowest-band eigenstates of $\hat{H}_{\text{eff}}$ and each of these $N_m$ states is occupied by $N_p$ atoms $|k, N_p\rangle =\mathcal{N} (a_k^{\dagger})^{N_p}|0\rangle$. 

A procedure for selecting the states $|k\rangle$ is described in Refs.~\cite{Dauphin2013, Aidelsburger2015}. 
In order to probe the Chern number of the lowest band, the states $|k\rangle$ should correspond closely to the lowest-band eigenstates of $\hat{H}_{\text{eff}}$. At the same time, in the experiment in the initial moment the atomic cloud is spatially localized. According to Refs.~\cite{Dauphin2013, Aidelsburger2015} the optimal approach is to consider a steep confining potential and to use the low-lying eigenstates of
\begin{equation}\label{eq:hini}
 \hat{H}_{\text{initial}} = \hat{h}^{\text{eff}}+\left(\frac{r}{r_0}\right)^\zeta,
\end{equation}
where in our calculations $\hat{h}^{\text{eff}}$ is either $\hat{H}_{\text{eff},0}$ from Eq.~(\ref{eq:heff0}) or $\hat{H}_{\text{eff},1}$ from Eq.~(\ref{eq:heff}) and the parameters of the confining potential are set to $r_0 = 20, \zeta = 20$.

The dynamics of the initial state (\ref{eq:rhoini}) is induced by a double quench: at $t_0=0$ the atomic cloud is released from the confining potential and exposed to a uniform force of intensity $F$ along the $y$ direction. During the whole procedure the driving providing the laser-assisted tunneling, defined in Eq.~(\ref{eq:drive}), is running.

The main observables of interest
are the center-of-mass position along $x$ direction
\begin{equation}
x(t) = \Big\langle \sum_{l,m} l\lvert\psi_{l,m}(t) \rvert^2\Big\rangle,
\end{equation}
and the population of the $i$th band of the effective model
\begin{equation}\label{eq:eta}
 \eta_i(t)=\Big\langle\sum_{|k\rangle\in i\text{-th band}}\Big\lvert \sum_{l,m} \alpha_{lm}^{k*} \psi_{lm}(t)\Big\lvert^2\Big\rangle,
\end{equation}
where the states $|k\rangle = \sum_{l,m} \alpha_{lm}^k|l, m\rangle$ correspond to the eigenstates of the effective model.
Here, angle brackets $\langle~\rangle$ denote averaging over $N_{\text{samples}}$  sets of initial conditions.

In the case of non-interacting particles, these and other quantities can be numerically accessed by solving the single-particle time-dependent Schr\"odinger equation for $N_m$ different initial states $|k\rangle$. 
This is equivalent to sampling the initial state according to Eq.~(\ref{eq:initial}).

In the end, we give two technical remarks.
First, all our calculations are done in the rotating frame; see Eq. \eqref{eq:htd} in Appendix \ref{a:heff}.
The staggered potential \eqref{eq:staggered} is removed in this way.
Second, in the case when the evolution is governed by the time-dependent Hamiltonian \eqref{eq:gp},
the initial state is multiplied by the operator $\mathrm{e}^{-i\hat{K}(0)}$
in order to properly compare these results to the ones obtained from the evolution governed by
the effective Hamiltonian \eqref{eq:gp2};
see Eq. \eqref{eq:floquet}.

\begin{figure}[!b]
\includegraphics[width=0.5\textwidth]{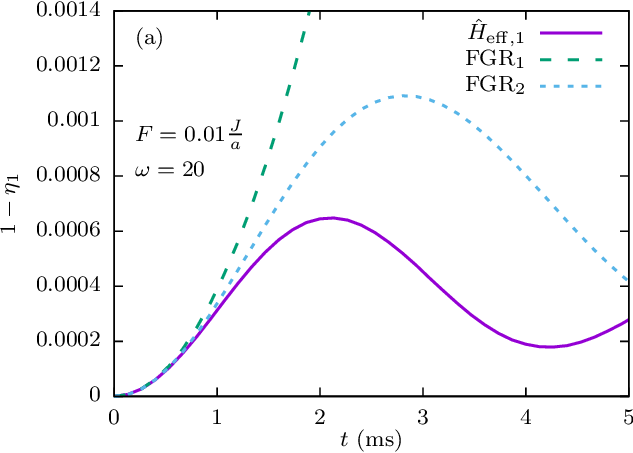}%
\includegraphics[width=0.5\textwidth]{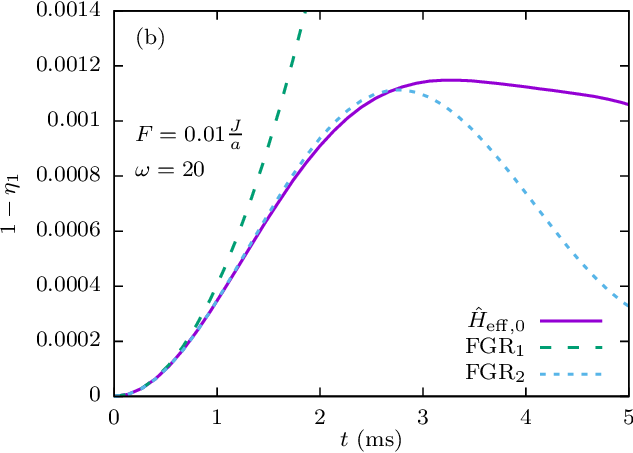}
\caption{Population in higher bands, comparison of numerical results (solid line) 
with the Fermi's golden rule in the first and second approximation (dashed lines).
Band populations are calculated for an initial BEC in an eigenstate of the effective Hamiltonian
and then averaged over (approximately) all states in the first band.
(a) Initial state and evolution from the effective Hamiltonian with correction $\hat{H}_\mathrm{eff,1}$, Eq. \eqref{eq:heff}.
(b) Without the correction, $\hat{H}_\mathrm{eff,0}$, Eq. \eqref{eq:heff0}.
\label{fig:fgr}}
\end{figure}

\section{Initial quadratic regime}\label{a:qz}
 
For simplicity, we will consider only the case without the confining potential and with very weak force $F=0.01$.
The initial state is a Bose-Einstein condensate in one of the eigenstates of the effective Hamiltonian. 
The results are later averaged over all first band eigenstates.

Fermi's golden rule predicts that the probability for transition from an initial state $\psi_i$
to a final state $\psi_f$, induced by a perturbation $\Delta\hat{H}$, is proportional to the square of matrix elements 
$\lvert\langle \psi_i\lvert \Delta\hat{H} \lvert\psi_f\rangle\lvert^2$. 
In this case, the perturbation is $\Delta\hat{H}=F\hat{y}$.
If we assume that the probability of a particle being in the initial state is always 
$P_i(t)=\lvert \psi_i(t)\lvert^2\approx 1$, Fermi's golden rule predicts \cite{Merzbacher}
\begin{equation}\label{eq:fgr1}
 P_{i\rightarrow f}^\mathrm{FGR_1}(t)=\frac{1}{\hbar^2}\lvert\langle \psi_i\lvert \Delta\hat{H} \lvert\psi_f\rangle\lvert^2 t^2.
\end{equation}
If we now also consider transitions from the other states to the initial state, 
but keep the assumption that the populations in other states are small $P_{j\neq i}(t)=\lvert \psi_{j\neq i}(t)\lvert^2\ll 1$, 
the time-dependent perturbation theory then predicts \cite{Merzbacher}
\begin{equation}\label{eq:fgr2}
 P_{i\rightarrow f}^\mathrm{FGR_2}(t)=\lvert\langle i\lvert\Delta\hat{H}\lvert f\rangle \lvert ^2 \frac{1-2\mathrm{e}^{-\frac{\Gamma}{2\hbar}t}\cos\left(\frac{E_f-E_i}{\hbar}t\right)+\mathrm{e}^{-\frac{\Gamma}{\hbar}t}}{\left(E_f-E_i\right)^2+\frac{\Gamma^2}{4}},
\end{equation}
where $\Gamma=\frac{2\pi}{\hbar}\lvert\langle i\lvert\Delta\hat{H}\lvert f\rangle \lvert ^2$
and $E_i$ ($E_f$) is the energy of the initial (final) state.

We plot the numerical results and both theoretical predictions from Fermi's golden rule in Fig.~\ref{fig:fgr}.
Here we can see that all three curves agree well for short times,
the second approximation longer remains close to the numerical results,
and that the initial quadratic regime is reproduced by theory.
This is the so-called quantum Zeno regime \cite{Debierre2015}.

\begin{figure}[!t]
\includegraphics[width=0.5\textwidth]{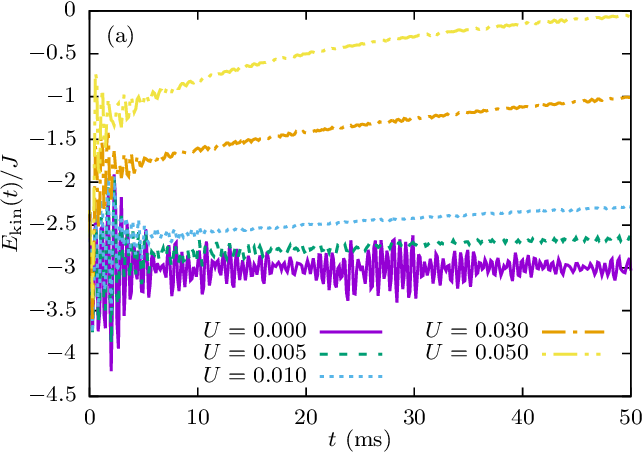}%
\includegraphics[width=0.5\textwidth]{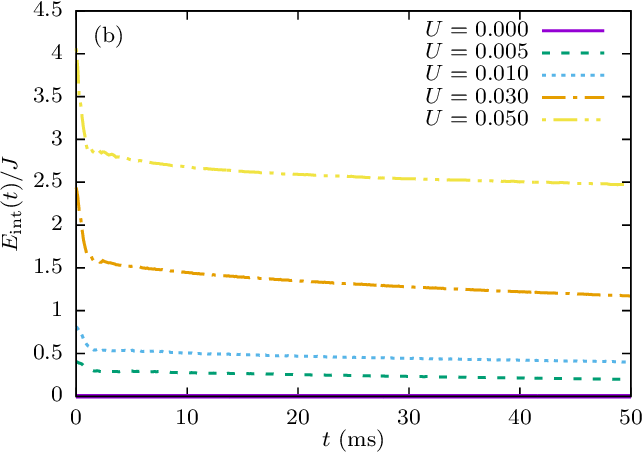}
\caption{(a) Kinetic energy per particle (expectation value of the time-dependent Hamiltonian 
$E_\mathrm{kin}(t)=\frac{1}{N}\Big\langle \sum_{l,m,i,j} \psi^*_{l,m}(t) H_{lm, ij}(t)\psi_{i,j}(t)\Big\rangle$ divided by the total number of particles $N$)
for several different interaction strengths.
(b) Interaction energy per particle $E_\mathrm{int}(t)=\frac{1}{N}\frac{U}{2}\Big\langle \sum_{l,m} \lvert\psi_{l,m}(t) \rvert^2\left( \lvert\psi_{l,m}(t) \rvert^2-1\right)\Big\rangle$.
$U$ is given in units where $J=1$.
\label{fig:energy}}
\end{figure}

\section{Energy}\label{a:en}

Time evolution of kinetic and interaction energy per particle for different interaction strengths 
is plotted in Fig.~\ref{fig:energy}.
Here we define the kinetic energy per particle as the expectation value of the time-dependent Hamiltonian \eqref{eq:htd}
divided by the total number of particles
$E_\mathrm{kin}(t)=\frac{1}{N}\Big\langle \sum_{l,m,i,j} \psi^*_{l,m}(t) H_{lm, ij}(t)\psi_{i,j}(t)\Big\rangle$,
while the interaction energy per particle is 
$E_\mathrm{int}(t)=\frac{1}{N}\frac{U}{2}\Big\langle \sum_{l,m} \lvert\psi_{l,m}(t) \rvert^2\left( \lvert\psi_{l,m}(t) \rvert^2-1\right)\Big\rangle$.
Both energies grow with increasing interaction coefficient $U$.

When the interactions are strong enough and after long enough time, 
the atoms become equally distributed between the eigenstates of the Hamiltonian $\hat{H}(t)$.
As the energy spectrum of $\hat{H}(t)$ is symmetric around zero,
the expectation value of $\hat{H}(t)$ (kinetic energy) should be zero when all bands are equally populated.
We can see this in Fig.~\ref{fig:energy}(a), where
the kinetic energy approaches zero at $t\approx50~\mathrm{ms}$ for the case $U=0.05$.

The interaction energy at first rapidly decreases, 
as the cloud rapidly expands after turning off the confinement potential $\hat{V}_\mathrm{conf}$, 
and after that continues to slowly decrease as the cloud slowly expands; 
see Fig.~\ref{fig:energy}(b).

These considerations also provide a possibility to discuss the applicability of the approximative method introduced in Sec. \ref{s:int}.
As we work in the regime of high frequency $\omega = 20$, we find that for weak interaction, at short enough times of propagation, the energy is approximately conserved. At stronger values of $U \geq 0.01$ we observe a slow increase in the total energy on the considered time scales.
In both cases we do not find the onset of parametric instabilities \cite{Lellouch2017}. If present, these instabilities are signaled by an order of magnitude increase in energy on a short time scale, that we do not find. 

\begin{figure}[!h]
\includegraphics[width=0.99\textwidth]{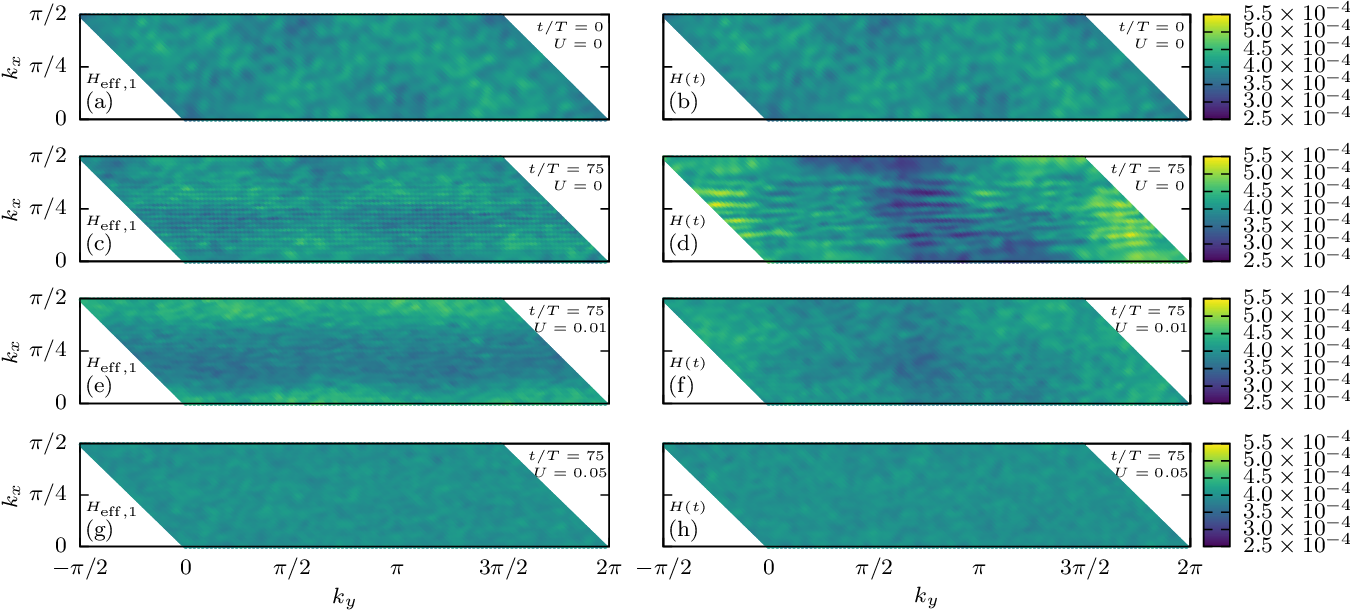}
\caption{Momentum-space density distribution in all bands, 
$\eta_1(\mathbf{k})+\eta_2(\mathbf{k})+\eta_3(\mathbf{k})$.
$U$ is given in units where $J=1$.
Left: evolution using the time-dependent Hamiltonian $\hat{H}_\mathrm{eff,1}$.
Right: evolution using the time-dependent Hamiltonian $\hat{H}(t)$.
(a), (b) Initial state.
(c), (d) Final state after $50~\mathrm{ms}$ ($75$ driving periods), noninteracting case $U=0$.
(e), (f) $U=0.01$.
(g), (h) $U=0.05$.
\label{fig:kdensity}}
\end{figure}

In addition, the two-body interaction can deplete the occupancies of initial coherent modes \cite{Choudhury2015, Bilitewski2015} and limit the validity of our approach. In principle, these types of processes can be addressed by including quantum fluctuations along the lines of the full truncated Wigner approach \cite{Polkovnikov2010}. Yet, we set our parameters in such a way that these additional contributions are small.

\section{Momentum-space density distribution}\label{a:kd}

The momentum-space probability densities 
at the initial moment and after $75$ driving periods ($50~\mathrm{ms}$)
are shown in Fig.~\ref{fig:kdensity}. 
The interactions deplete the lowest band, but also smooth out the density distribution.

\end{widetext}

\newpage
%

 \end{document}